\documentclass[conference]{style/IEEEtran}

\makeatletter
\providecommand*\input@path{}
\g@addto@macro\input@path{{style/}}
\makeatother

\usepackage{graphicx}
\usepackage{fontawesome5}
\usepackage{marvosym}
\usepackage{tikz}
\usetikzlibrary{arrows.meta, positioning, shapes.geometric, calc, fit, backgrounds}
\usepackage{amsmath}
\usepackage{amssymb}
\usepackage{cite}
\usepackage{subcaption}
\usepackage{textcomp}
\usepackage[dvipsnames,table]{xcolor}
\usepackage{booktabs}
\usepackage{multirow}
\usepackage{longtable}
\usepackage{xspace}
\usepackage{balance}
\usepackage{threeparttable}
\usepackage{url}
\usepackage{hyperref}
\usepackage{cleveref}
\usepackage[most]{tcolorbox}

\newcommand{\citeb}{{\textsc{CiteVerifier}}\xspace}

\crefname{section}{Section}{Sections}
\crefname{figure}{Figure}{Figures}
\crefname{table}{Table}{Tables}
\crefname{equation}{Equation}{equations}
\crefname{algocf}{Algorithm}{algorithms}
\crefname{appendix}{Appendix}{appendices}

\definecolor{paperPrimary}{HTML}{8888EE}
\definecolor{paperPrimaryLight}{HTML}{FAFAFF}
\definecolor{paperAlert}{HTML}{C75B39}
\definecolor{paperAlertLight}{HTML}{FFF0E6}
\definecolor{paperPositive}{HTML}{2E7D32}

\definecolor{tableRowLight}{gray}{0.95}
\definecolor{tableRowWhite}{gray}{1.0}
\definecolor{colGroupBlue}{HTML}{E8F4F8}
\definecolor{colGroupOrange}{HTML}{FFF0E6}
\definecolor{legendBlue}{HTML}{2E7D9A}
\definecolor{legendOrange}{HTML}{C75B39}

\definecolor{NKU}{HTML}{711a5f}
\definecolor{GiantsOrange}{HTML}{fe5a1d}

\newtcolorbox{keyfindingsSidebar}[1]{
	enhanced,
	colback=white,
	colframe=paperPrimaryLight,
	boxrule=0pt,
	arc=0pt,
	left=8pt,
	right=8pt,
	top=8pt,
	bottom=8pt,
	borderline west={4pt}{0pt}{paperPrimary},
	coltitle=paperPrimary,
	fonttitle=\bfseries\small,
	title={#1},
	before skip=8pt,
	after skip=8pt
}

\newtcolorbox{casestudybox}[1]{
	enhanced,
	colback=white,
	colframe=black!86,
	boxrule=1pt,
	arc=3pt,
	left=2pt,
	right=8pt,
	top=4pt,
	bottom=3pt,
	fonttitle=\bfseries\small,
	title={#1},
	colbacktitle=black!60,
	coltitle=white!55,
	before skip=8pt,
	after skip=8pt
}

\newcommand{\mkStar}{\textcolor{blue}{*}}
\newcommand{\mkDag}{\textcolor{red}{\textdagger}}
\newcommand{\mkDDag}{\textcolor{cyan}{\textdaggerdbl}}
\newcommand{\mkEmail}{\textcolor{green!70!black}{\Letter}}

\begin{document}

\title{\textsc{GhostCite}: A Large-Scale Analysis of Citation Validity in the Age of Large Language Models}

\author{
	\IEEEauthorblockN{Zuyao Xu\textsuperscript{\mkDag\mkStar}, Yuqi Qiu\textsuperscript{\mkDag\mkStar}, Lu Sun\textsuperscript{\mkDag\mkStar}, Fasheng Miao\textsuperscript{\mkDDag}, Fubin Wu\textsuperscript{\mkDag}, Xiang Li\textsuperscript{\mkDag\mkEmail}}
	\IEEEauthorblockN{Xinyi Wang\textsuperscript{\mkDag}, Haozhe Lu\textsuperscript{\mkDag}, Zhengze Zhang\textsuperscript{\mkDag}, Yuxin Hu\textsuperscript{\mkDag}, Jialu Li\textsuperscript{\mkDag}, Luo Jin\textsuperscript{\mkDag}}
	\IEEEauthorblockN{Feng Zhang\textsuperscript{\mkDDag}, Rui Luo\textsuperscript{\mkDag}, Xinran Liu\textsuperscript{\mkDag}, Yingxian Li\textsuperscript{\mkDag}, Jiaji Liu\textsuperscript{\mkDag}}
	\IEEEauthorblockA{\textsuperscript{\mkDag}Nankai University\quad \textsuperscript{\mkDDag}Tsinghua University\\
	\mkEmail\ \texttt{lixiang@nankai.edu.cn}\\
	\textit{\mkStar\ Equal contribution.}\;\textit{\mkEmail\ Corresponding author.}}
}

\maketitle

\begin{abstract}
Citations provide the basis for trusting scientific claims; when they are invalid or fabricated, this trust collapses.
With the advent of Large Language Models (LLMs), this risk has intensified: LLMs are increasingly used for academic writing, but their tendency to fabricate citations (``ghost citations'') poses a systemic threat to citation validity.
To quantify this threat, we develop \citeb, an open-source framework for large-scale citation verification, and conduct a comprehensive study of citation validity in the LLM era through three complementary experiments.
First, we benchmark 13 LLMs on citation generation task in various research domains, finding that all models hallucinate citations at rate from 14.23\% to 94.93\%.
Second, we analyze 2.2 million citations from 56,381 papers at AI/ML and Security venues (2020--2025), finding that 1.07\% of papers contain invalid citations, with an 80.9\% increase in 2025.
Third, we survey 97 researchers, finding that 87.2\% use AI-powered tools in their workflows, 76.7\% of reviewers do not thoroughly check references, and 74.5\% view peer review as ineffective at catching citation errors.
Based on these findings, we argue that ghost citations represent a systemic threat to academic integrity, and call for coordinated efforts from community to address this challenge.
\end{abstract}

\IEEEpeerreviewmaketitle

\section{Introduction}

Citations serve as a trust mechanism that justifies scientific claims~\cite{kaplan1965norms}: readers assume that cited work exists and supports the claims being made, and rarely verify that it does~\cite{simkin2003read}. However, when a citation is invalid, this trust mechanism collapses.
A fabricated citation can misattribute foundational ideas, introduce phantom prior work, misdirect subsequent research investment, or undermine the evidentiary chain of a claimed contribution, threatening the integrity of the entire research community.
The advent of AI-assisted writing has introduced a novel threat to this academic trust infrastructure.
Researchers increasingly leverage LLMs to synthesize literature reviews, curate related work, or compile reference lists~\cite{mohammadi2025generative}.
When prompted for citations, these systems do not look up references. Instead, leveraging their inherent generative nature, they \textit{synthesize} references by combining statistically correlated tokens (real author names, plausible-sounding titles, and prestigious venue names) into citations that appear authentic despite being entirely fabricated~\cite{vaswani2017attention}. 
This phenomenon commonly described as ``ghost citations''~\cite{consultmu2026ghost,tay2025why}, representing a new category of academic misconduct: high-fidelity fabrications that exploit the format-compliance of bibliographic conventions to evade detection. 
The emergence of ghost citations has caught the attention of conferences and governing bodies in academia. For instance, ICLR 2026 chairs issued guidelines warning against AI-generated content, including fabricated citations~\cite{chairs2025iclr}, and major publishers such as IEEE have established policies requiring disclosure of AI usage in manuscripts~\cite{ieee2025ai-policy}.
Recent reports also document hallucinated citations in conference submissions and accepted papers, as well as in domain-specific analyses~\cite{gptzero2025nips,gptzero2026iclr,sakai2026hallucitation}.
However, despite growing awareness, we still lack systematic answers to three fundamental questions: 
\begin{itemize}
    \item \textit{Q1: How frequently do LLMs hallucinate citations across research domains?} 
    \item \textit{Q2: To what extent have invalid citations entered published academic literature?} 
    \item \textit{Q3: Why do authors and reviewers fail to detect them?} 
\end{itemize}

\noindent
\textbf{Research Gap}
Prior work and policy discussions highlight the risks of ghost citations~\cite{chairs2025iclr,thorp2023chatgpt,oladokun2025hallucitation}, but to answer the above questions, three practical gaps persist.
First, the scale of the problem is unknown: we have no empirical measurement of how often LLMs fabricate and how many invalid citations have entered the published literature.
Second, there is no scalable way to detect citations validality, because they often deviate from standard formats and need to be verified against multiple bibliographic sources.
Third, we lack a clear understanding of how invalid citations go through reseacher's draft and peer review process and end up in the published record without being checked.
These gaps motivate our work, which develops a technical framework for large-scale citation verification and applies it to systematically analyze the prevalence of ghost citations in LLM outputs and the published record, as well as the human behaviors that allow them to persist.

\noindent
\textbf{Our Study.}
In this paper, we develop \citeb, 
an open-source framework for large-scale citation verification, and conduct three complementary experiments that answer the above research questions.

\textit{Experiment 1: LLM Benchmark}
We evaluate 13 state-of-the-art LLMs across 40 computer science research domains aligned with arXiv CS subject classes~\cite{arxivCSClasses}, finding that all models hallucinate citations at rates ranging from 14.23\% to 94.93\%, with substantial variation across domains (Section~\ref{sec:llm_results}).

\textit{Experiment 2: Archival Analysis.}
We collected~\cite{bailey2012menlo} 56,381 papers from eight AI/ML and Security venues spanning 2020 to 2025. After employing \citeb to analyze 2.2 million citations of these papers, our system flagged 2,530 citations for unmathced metadata.
We then manually verified each one, identifying 739 invalid citations across 604 papers (1.07\% of 56,381) are definitively invalid, with an 80.9\% increase in invalid citation rates in 2025 (from a 2020--2024 average of 0.89\% to 1.61\%), and observed phenomena of error propagation across papers (Section~\ref{sec:archival_results}).

\textit{Experiment 3: User Study.}
We issued a survey to 97 researchers across various roles  and research domains, asking about their practices and perceptions regarding citation validity, and received 94 valid responses. We find 41.5\% of authors admit to copy-pasting BibTeX entries without checking. 76.7\% of reviewers do not thoroughly inspect references, 74.5\% of all respondents view peer review as ineffective at catching citation errors, while 70.2\% strongly support the idea of automated checks to catch citation errors (Section~\ref{sec:survey_results}).

Our findings reveal that unreliable AI tools, combined with inadequate human verification, facilitate the penetration of fabricated citations into published literature. Based on our empirical evidence, We discuss implications and propose interventions for researchers, venues, and tool developers, highlighting the need for coordinated efforts to address this emerging threat to academic integrity.
Our work follows the tradition of meta-science measurement studies that expose structural problems in research practices~\cite{baker2016reproducibility,ioannidis2005most,rossow2012prudent,schloegel2025confusing}.
Just as prior work revealed reproducibility crises and evaluation pitfalls~\cite{baker2016reproducibility,rossow2012prudent}, we identify an emerging threat to the integrity of academic citation, and provide a technical framework and empirical evidence to inform the community's response to this challenge.

\noindent
\textbf{Contributions.}
Our contributions are as follows:

\begin{enumerate}
    \item \textbf{Technical baseline for citation verification.} We develop \textsc{CiteVerifier}, an open-source framework that verifies citations at scale, providing a replicable implementation example for researchers and future work.

    \item \textbf{Comprehensive benchmark of LLM citation hallucination.} We evaluate \textit{13} LLMs across \textit{40} domains on \textit{375,440} generated citations, revealing widespread citation hallucination across models and domains.
    
    \item \textbf{Empirical analysis of published literature.} We analyze \textit{2.2M} citations from \textit{56,381} papers (2020--2025), documenting the presence and increasing prevalence of invalid citations in the published literature.
    
    \item \textbf{User study of researcher practices.} We survey \textit{97} researchers across various roles and research domains, characterizing AI adoption rates, verification behaviors, and community attitudes toward citation validity and potential interventions.
 
\end{enumerate}

\section{Background and Related Work}
\label{sec:background}

Citations serve as a trust mechanism that justifies scientific claims; the validity of citations is therefore central to the integrity of scholarly communication~\cite{greenberg2009how}.
When citations are invalid, this trust breaks down: readers may be misled by references that do not support, or do not even exist to support, the claims being made.
Such fabrications are commonly termed ``ghost citations''~\cite{consultmu2026ghost,tay2025why}.
To understand and address this threat, we first need to clarify three fundamental questions: What exactly are ghost citations, how did they arise, and why do they pose such a risk to scientific research?

\subsection{Ghost Citations in the Wild}
\label{sec:background:ghost}
Citation errors existed long before generative AI became widespread.
Early citation-analysis studies reported that citation errors, including incorrect author names, publication years, page numbers, and even entirely non-existent sources, have been a persistent problem in academic publishing~\cite{sweetland1989errors}.
Furthermore, Simkin and Roychowdhury~\cite{simkin2005stochastic} estimated that a substantial fraction of citations are copied from secondary sources without consulting the original work, propagating errors through the literature.
These human errors, while problematic, typically arose from carelessness or negligence rather than intentional fabrication.

However, the emergence of LLMs fundamentally transformed the nature of citation errors, introducing a fundamentally different phenomenon: the systematic fabrication of references that never existed\cite{oladokun2025hallucitation}.
Unlike human mistakes, LLM-generated citations are systematically produced through autoregressive token prediction~\cite{vaswani2017attention}.
When generating academic text, LLMs identify high-probability token clusters associated with a research field (authors like ``Vaswani,'' venues like ``NeurIPS,'' and domain-specific terminology) and combine them into plausible but non-existent citations.
As Bender et al.~\cite{bender2021dangers} noted, LLMs function as stochastic parrots that prioritize the structure of language over its truth.
Because academic citations follow rigid syntactic templates, LLMs can effortlessly mimic this surface form, producing fabricated references indistinguishable from authentic ones.

The phenomenon of LLM hallucination has been extensively studied~\cite{ji2023survey,zhang2025siren,huang2023survey}, with factuality benchmarks such as TruthfulQA~\cite{lin2022truthfulqa} and HaluEval~\cite{li2023halueval} quantifying these failures.
As generative AI tools became widespread in academic contexts~\cite{lund2023chatgpt,cotton2024chatting,dwivedi2023so}, reports of fabricated citations accumulated in academic blogs~\cite{consultmu2026ghost,tay2025why}.
Detection efforts have emerged: GPTZero reported hallucinated citations in both ICLR 2026 submissions and NeurIPS 2025 accepted papers~\cite{gptzero2026iclr,gptzero2025nips}, while watermarking~\cite{kirchenbauer2023watermark} and zero-shot detection based on probability curvature~\cite{mitchell2023detectgpt} offer complementary strategies.
Nevertheless, these tools primarily target AI-generated prose rather than the factual accuracy of citations themselves. Recent reports and empirical studies further illustrate the problem's scope; ACL-focused analyses identified hundreds of hallucinated references~\cite{sakai2026hallucitation}, and studies in librarianship documented hallucinated citations in ChatGPT outputs~\cite{oladokun2025hallucitation}.
Nevertheless, we still lack systematic, large-scale analysis of ghost citation prevalence across models and the published literature, as well as a clear understanding of why verification
behaviors fail to prevent their propagation.

\subsection{The Risk to Scientific Research}
\label{sec:background:record}

The spread of ghost citations goes beyond mere typographical or formatting errors; it constitutes a systemic threat to the integrity of scholarly communication, with the capacity to undermine scientific research at scale. We analyze these risks across three key stakeholder groups.

\noindent
\textbf{Impact on Researchers.}
In normal scholarship, citations ground claims in verifiable prior work and help researchers navigate the literature with confidence~\cite{kaplan1965norms,merton1973sociology}.
Ghost citations undermine this process by inserting fabricated references that appear legitimate, misdirecting research efforts and wasting resources on phantom threads; when they enter literature reviews or are used as starting points for new investigations, they corrupt scientific claims and propagate misinformation through the research pipeline.
In this situation, ghost citations create \textit{Epistemic Pollution}: non-existent studies are cited to support claims, creating an illusion of evidential support~\cite{greenberg2009how}.

\noindent
\textbf{Impact on Reviewers and Publishers.}
The research community operates on a foundational presumption of trust: we assume authors work in good faith, and reviewers rarely verify the work exhaustively~\cite{smith2006peer}.
Plausible-looking LLM citations exploit this trust gap; verifying 30 to 50 references per paper would place unrealistic demands on the already-strained peer review process.
Our own manual verification required a 16-person research team working for about one month to review 2,530 flagged citations (\Cref{sec:archival_results}), demonstrating that thorough checking demands substantial human time and coordination.
Publishers and conference organizers lack automated tools to detect fabricated citations, so human reviewers need to validate references manually, a task that is time-consuming and difficult to perform systematically.

\noindent
\textbf{Impact on the Research Community.}
Research impact analysis relies on the citation graph (nodes = publications, edges = citations)~\cite{newman2001structure}.
Ghost citations inject phantom nodes and invalid edges, undermining citation-based indicators (e.g., Impact Factor, h-index) and network metrics (co-citation, bibliographic coupling), degrading the reliability of quantitative impact assessment.
More fundamentally, as fabricated citations accumulate in the published record, they erode the trust infrastructure that supports academic communication.
If future LLMs are trained on papers containing ghost citations, these fabrications may be reproduced amplifying errors across generations~\cite{shumailov2024ai}.
Ultimately, widespread ghost citations may force the research community to shift from presumptive trust to systematic skepticism, imposing unsustainable verification burdens on scientific communication.

\begin{figure*}[t]
    \centering
    \includegraphics[width=0.98\textwidth]{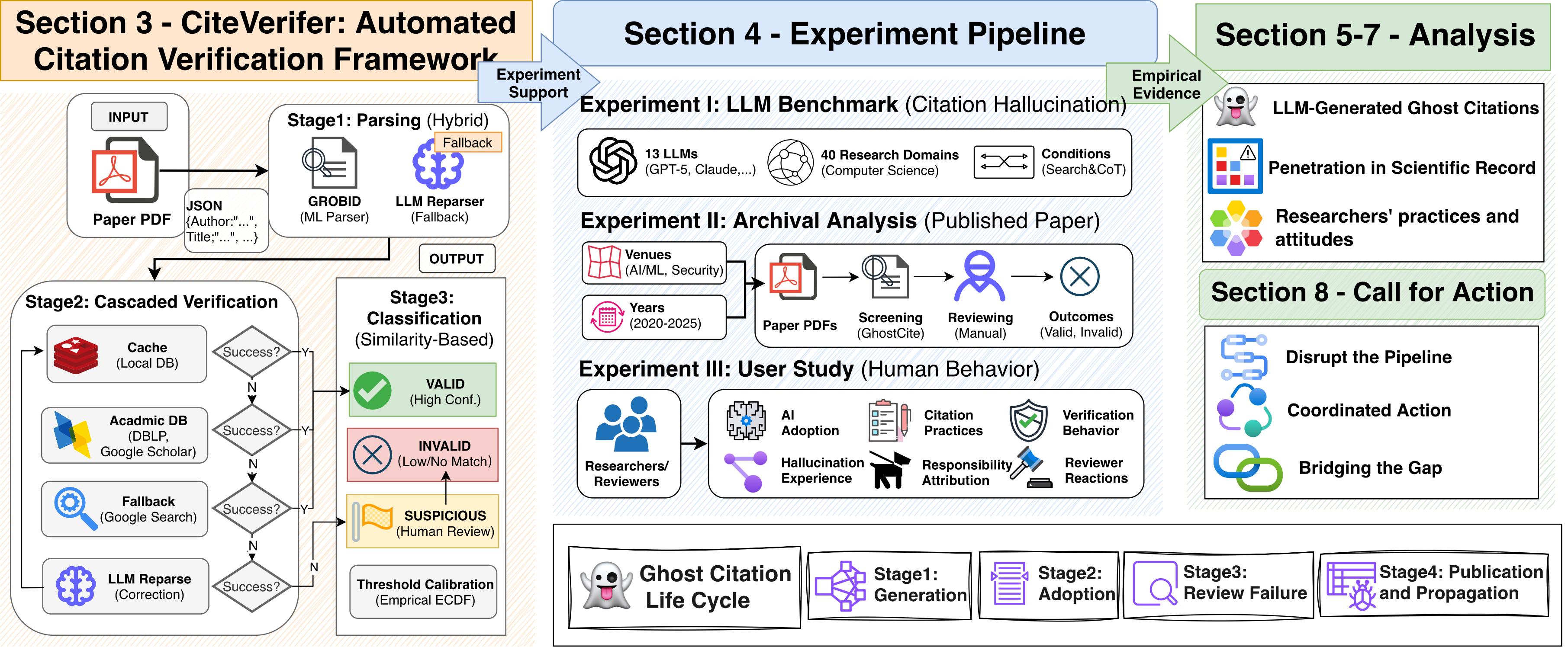}
    \caption{The framework of \citeb and experimental pipeline.}
    \label{fig:methodology_overview}
\end{figure*}

\subsection{Research Questions}
\label{sec:research_gap}

Recent reports and targeted analyses have documented hallucinated citations in conference submissions and accepted papers~\cite{gptzero2026iclr,gptzero2025nips}, as well as in ACL-focused studies~\cite{sakai2026hallucitation}. However, despite this growing evidence, three critical questions remain unanswered.

\noindent
\textbf{RQ1: How can ghost citations be detected at scale?}
Unlike AI-generated prose detection, citation verification requires cross-referencing against external bibliographic databases, a challenge complicated by the vast heterogeneity of citation formats~\cite{ISO690_2021}, the incompleteness of existing academic indices, and the absence of standardized verification protocols.
Major publishers including Nature, Science, and IEEE have issued policies requiring disclosure of AI usage~\cite{nature2023tools,thorp2023chatgpt,ieee2025ai-policy}, and academic conferences such as ICLR 2026 have established explicit guidelines~\cite{chairs2025iclr}.
However, policy frameworks remain fragmented, and no scalable, automated verification infrastructure exists to systematically detect fabricated references across the published literature.

\noindent
\textbf{RQ2: What is the current prevalence of ghost citations in the published literature?}
Prior meta-science measurement studies have demonstrated the value of large-scale empirical analysis in exposing systemic issues, from reproducibility crises~\cite{baker2016reproducibility,ioannidis2005most} to evaluation pitfalls in machine learning~\cite{lipton2018troubling,wagstaff2012machine} and citation manipulation practices~\cite{fister2016toward}.
However, no comparable study has systematically quantified the prevalence and characteristics of ghost citations.
Existing evidence remains largely informal, lacking the rigorous methodology that is necessary to determine whether ghost citations represent just an isolated incident or a widespread undermining of the whole scientific research.

\noindent
\textbf{RQ3: Why do verification practices fail to prevent ghost citations from reaching publication?}
Surveys and guidelines routinely advise researchers to verify AI-generated contents\cite{cotton2024chatting,dwivedi2023so}, and many researchers report that they do so\cite{lee2025impact}.
But whether researchers' self-reported verification behaviors match their actual practices has not been systematically studied: it remains unclear how often researchers rely on AI-generated citations, under what conditions verification is skipped, and which behavioral or cognitive factors allow fabricated citations to persist in the pipeline from model output to published paper.

Our work aims to address these three research questions through a comprehensive, multi-dimensional analysis of ghost citations in the LLM era, combining technical development with empirical measurement and behavioral analysis.
We develop \citeb, an automated citation verification pipeline with robust parsing, cascaded multi-source retrieval, and calibrated similarity matching to achieve the high accuracy and support high-throughput validation. Leveraging \citeb, we conduct three complementary studies: (1) a systematic benchmark of citation hallucination rates across 13 LLMs under varying conditions (RG1; \Cref{sec:llm_results}); (2) a large-scale audit of 56,000+ published papers to quantify real-world prevalence (RG2; \Cref{sec:archival_results}); and (3) a user study investigating the behavioral factors that enable fabricated citations to reach publication (RG3; \Cref{sec:survey_results}).

\section{Automated Verification Framework}
\label{sec:methodology}

To systematically analyze the prevalence and impact of ghost citations, we first need a reliable method to identify them at scale.
To this end, we developed \citeb, an automated citation verification framework designed to robustly identify invalid citations across large corpora of academic papers.
In this section, we describe the design and implementation of our \citeb framework, detailing how it addresses the significant challenges of citation verification to enable our subsequent analyses.


\subsection{Framework Design}
\label{subsec:framework_design}

\noindent
\textbf{Design Challenges.}
Verifying citations at scale raises several technical and practical challenges; we summarize these challenges below, then state the design goals we set to address them.
\begin{itemize}
    \item \textit{Parsing heterogeneity}: academic citations exhibit enormous format diversity, varying across venues, disciplines, and even individual papers; PDF extraction introduces additional noise through OCR errors, non-standard fonts, so a robust pipeline needs to tolerate parsing imperfections without generating excessive false positives.
    \item \textit{Database coverage}: no single bibliographic database provides comprehensive coverage; different sources have different profiles, so absent results are often ambiguous, motivating a multi-source approach.
    \item \textit{Verification scalability}: verifying millions of citations requires a well-engineered framework design, such as rate limiting, cache reuse, and concurrent requests, to achieve practical throughput without excessive costs.
\end{itemize}

We address these challenges by designing \citeb as a modular, cascaded pipeline, as illustrated in \Cref{fig:methodology_overview}.
It parses reference strings into structured metadata, queries sources in order of cost and precision, and classifies validity using similarity-based matching to tolerate minor variations.
An LLM-based reparser serves as a last-resort fallback for malformed inputs.
To balance coverage and precision, it catches fabricated references while minimizing false positives.

\subsection{Framework Implementation}
\label{subsec:framework_impl}

We implement \citeb as a three-stage pipeline: \textit{parsing}, \textit{verification}, and \textit{classification}, as detailed below.

\noindent
\textbf{Stage 1: Reference Parsing.}
The parsing stage extracts structured metadata from raw citation strings.
We use GROBID~\cite{grobid-client-python} as the primary parser; when initial parsing fails, a Qwen3-Flash~\cite{bai2023qwen} based LLM reparser extracts metadata via a JSON schema and re-runs the verification chain.

\noindent\textbf{Stage 2: Cascaded Verification.}
With structured metadata extracted, the verification stage checks citation validity using four verification steps executed sequentially:

\begin{enumerate}
    \item \textit{Local Cache Check.}
    We first query a local SQLite database of previously verified citations.
    This cache stores results from prior searches, enabling efficient re-verification and reducing API costs.
    The cache is keyed by normalized title and returns stored verification results along with matched external reference metadata.
    
    \item \textit{Academic Database Query.}
    For uncached entries, we query multiple bibliographic databases in cascade.
    We construct a local DBLP database from the XML dump for efficient local queries, and access Google Scholar data through the ScrapingDog API proxy~\cite{scrapingdog} for recent or non-indexed papers.
    The query is constructed from the citation title, and returned results are parsed to extract title, authors, year, and venue metadata. We store all results in the cache for future queries.

    \item \textit{Web Search Fallback.}
    If the academic databases fail to find the reference, we fall back to general web search.
    This broader search covers recent papers not yet indexed by academic databases, technical reports, preprints, and non-traditional academic content.

    \item \textit{LLM Reparse Fallback.}
    If all prior steps fail to find a match, we hypothesize that parsing errors may have corrupted the search query.
    The system invokes the LLM reparser to re-extract metadata from the raw string, then re-executes the verification chain with the corrected metadata.
    This addresses cases where minor parsing errors, ensuring accuracy even with noisy inputs.
\end{enumerate}

Each step returns a result if verification succeeds or proceeds to the next step until all options are exhausted, at which point the citation is classified as invalid.

\noindent\textbf{Stage 3: Similarity-Based Classification.}
After retrieving candidate matches from the verification stage, we classify the citation as \texttt{Valid} or \texttt{Invalid} based on the similarity between the input citation and retrieved candidates.
In this part, we only focus on title similarity, as our preliminary goal is to identify citations that are untraceable or even non-existent (i.e. ghost citations), rather than citation errors as reported in prior work~\cite{simkin2005stochastic, simkin2003read},
The classfication is based on the highest similarity score among all retrieved candidates, using a empirical calibrated threshold $\theta$ to determine validity: if the maximum similarity exceeds $\theta$, we classify the citation as \texttt{Valid}; otherwise, it is classified as \texttt{Invalid}.

\subsection{Implementation Details}
We implement \citeb in Python with asynchronous I/O (asyncio) for concurrent API requests, enabling efficient parallel processing of multiple citations.

\begin{figure}[t]
    \centering
    \includegraphics[width=\columnwidth]{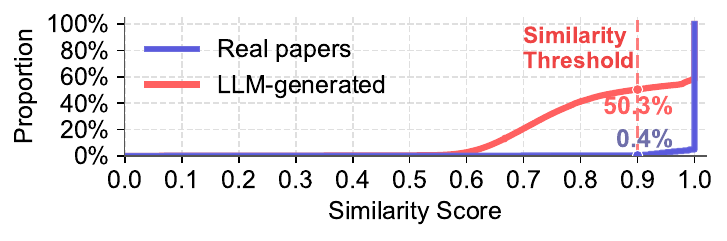}
    \vspace{-5.5mm}
\caption{ECDF of title similarity scores: LLM-generated vs.\ real-paper citations.}
    \label{fig:llm_papers_similarity_ecdf}
\end{figure}

\noindent\textbf{Similarity Computation and Threshold Selection.}
For similarity-based classification, we normalize titles and compute Levenshtein distance\cite{levenshtein1966binary} (using the highest similarity among all retrieved candidates).
We set the classification threshold to $\theta = 0.9$, roughly corresponding to a one or two word difference in a long title.
We validate this choice based on the subsequent experiments in \cref{sec:llm_results} and \cref{sec:archival_results}, using the empirical CDF of title similarity for real-paper citations and LLM-generated citations (\Cref{fig:llm_papers_similarity_ecdf}).
Real-paper similarities remain near 1.0: only 0.4\% fall at or below 0.9, and the small tail below 1.0 is largely attributable to noise (OCR errors, hyphenation, minor typos).
In contrast, LLM-generated citations rise gradually, with 50.3\% at or below 0.9, validating $\theta = 0.9$ as a threshold that preserves high recall for real papers while filtering low-similarity hallucinations.

\noindent\textbf{Practical Considerations.}
We use semaphore-based rate limiting (10 concurrent requests by default) to respect API quotas, batch-process entire directories with progress tracking and incremental export, and output results as CSV with columns for status, diagnosis, similarity scores, and matched metadata.
These choices aim to balance throughput, cost, and usability for large-scale citation verification.



\section{Experiment Design}
\label{subsec:experiment_design}

In this section, we describe the design of our three complementary experiments to analyze ghost citations in the LLM era. Each experiment targets a specific research question: the LLM Benchmark quantifies how often LLMs hallucinate citations and what factors affect their accuracy; the Archival Analysis measures the real-world prevalence of invalid citations in published papers; and the User Study investigates researchers' practices and attitudes toward citation verification. 
Below, we detail the setup, conditions, and rationale for each experiment.

\subsection{Experiment I: LLM Benchmark}
\label{subsec:experiment_i}

In this experiment, we systematically evaluate 13 state-of-the-art LLMs spanning major vendors and diverse architectural approaches, assessing their ability to generate valid academic citations.

\begin{table}[t]
  \centering
  \begin{threeparttable}
    \caption{Large language models evaluated in citation hallucination benchmark (13 models across major vendors).}
    \label{tab:models_evaluated}
    \small
\rowcolors{2}{tableRowLight}{tableRowWhite}
\begin{tabular}{lll}
\toprule
\textbf{Model} & \textbf{Vendor} & \textbf{Release Date} \\
\midrule
GPT-5 & OpenAI & 2025-08-07 \\
Claude-Sonnet-4 & Anthropic & 2025-05-22 \\
Gemini-2.5-Pro & Google DeepMind & 2025-06-17 \\
Grok-4-Fast & XAI & 2025-09-19 \\
Llama-4-Maverick & Meta & 2025-04-05 \\
Phi-4 & Microsoft & 2024-12-12 \\
GLM-4.5 & Zhipu AI & 2025-07-28 \\
Hunyuan-A13B-Instruct & Tencent & 2025-06-27 \\
UI-TARS-1.5-7B & ByteDance Seed & 2025-04-17 \\
Qwen3-Max & Alibaba Qwen & 2025-09-23 \\
DeepSeek-Chat-V3.1 & Deepseek & 2025-08-21 \\
ERNIE-4.5-21B-A3B & Baidu & 2025-06-30 \\
Kimi-K2-0905 & Moonshot & 2025-09-05 \\
\bottomrule
\end{tabular}
  \end{threeparttable}
\end{table}

\noindent\textbf{Setup.}
All models were accessed via a third-party API aggregation platform (OpenRouter)~\cite{openrouter} for consistent interfaces.
We constructed a taxonomy of 40 computer science research domains based on arXiv CS subject classes~\cite{arxivCSClasses} (full list in the \cref{app:domain_codes}).
For each model-domain combination we use a standardized prompt to generate citations within that domain; the prompt asks the model to produce a fixed number of domain-specific references in a strict JSON schema (author, title, venue, year, type), enabling consistent parsing and verification (the full prompt is provided in \cref{app:llm_citation_prompt}).

\noindent\textbf{Conditions and verification.}
We varied two factors: batch size (10, 20, or 30 citations per prompt) and whether to enable online search with chain-of-thought prompting.
For each model-domain-batch-search combination, we generate 120 citations ($30\times4$ interactions, $20\times6$ interactions, or $10\times12$ interactions).
In total, we obtained 375,440 citations from 22,800 API interactions, with total API costs amounting to $\approx 800$ USD.
All generated citations were verified with \citeb and classified as VALID or INVALID. Unparseable outputs were flagged as format errors and only included in format compliance statistics, not hallucination rates.

\noindent\textbf{Design rationale.} Our benchmark prompts models to generate fixed-size blocks of domain-specific citations. This approach ensures fair cross-model comparison under identical conditions, though it may differ from real-world use where citations are produced while drafting specific arguments. Accordingly, our hallucination rates should be interpreted as a controlled baseline rather than an estimate of real-world prevalence; we discuss interpretation further in Section~\ref{sec:discussion}.

\noindent\textbf{LLMs as Citation Validators.}
To assess whether LLMs can reliably evaluate citation validity, we conducted an auxiliary experiment where each model was prompted to verify 100 bibliographic entries with known ground truth (50 valid, 50 invalid) from our archival analysis (\cref{sec:archival_results}), one entry per prompt, and classify each as VALID or INVALID. The prompt is provided in Appendix~\ref{app:llm_judge_prompt}.

\subsection{Experiment II: Archival Analysis}
\label{subsec:experiment_ii}

To quantify the real-world penetration of ghost citation into the scientific record, we analyze published papers from top-tier venues.
We target two research communities: \textit{Security} (IEEE S\&P, USENIX Security, ACM CCS, NDSS) and \textit{AI/ML} (NeurIPS, ICML, IJCAI, AAAI), focusing on venues that are central to LLM research and have rigorous peer review~\cite{csrankings}.
Data span 2020--2025 (pre-LLM 2020--2022, post-LLM 2023--2025)~\cite{maslej2025artificial} for longitudinal analysis.
We collected all accepted papers (main, workshop, short, poster) in PDF from official proceedings using ethically compliant access and stewardship practices (see \cref{sec:ethics}), yielding a total of 56,381 papers.
To ensure the accuracy of our analysis and to avoid false positives (where valid citations are misclassified as invalid), we employ a rigorous three-stage process:
\begin{itemize}
    \item \textit{Step 1: Automated Examination.}
    All 56,381 papers were processed through \citeb, extracting citations, and flagging citations with similarity scores below $\theta = 0.9$ as potentially invalid for manual review. 
   \item \textit{Step 2: Manual Verification.} Sixteen trained research assistants manually reviewed all flagged citations over approximately one month. Each citation was independently checked at least twice before being classified into one of three categories: \textit{Non-Academic Source} (websites, blogs, repositories), \textit{Valid} (verified through extensive manual search), or \textit{Invalid} (incorrect metadata or untraceable).

    \item \textit{Step 3: False Negative Estimation.}
    To ensure accuracy, we sampled 400 citations from the ``valid'' pool for manual review (95\% confidence, 5\% margin of error).
    No additional invalid citations were found.
\end{itemize}

Two researchers subsequently reviewed all manually classified invalid citations to ensure accuracy, and further subdivided them into two subcategories: \textit{error citation} and \textit{ghost citation} (untraceable citations cannot be found with extensive manual search).
Some untraceable citations appear clearly fabricated (e.g., titles stitched together from high-probability phrases); others simply do not exist in standard bibliographic indexes, both align with the definition of ``ghost citations'' in our study.

\subsection{Experiment III: User Study}
\label{subsec:experiment_iii}

To understand the human factors enabling ghost citations to reach publication, we conducted a survey of active researchers across career stages and research areas, recruiting participants via public accsessable social media and targeted emails to 300 randomly sampled authors and PC members from the above 8 venues;  
The survey covered four dimensions: AI adoption, citation practices, author-level verification behavior, and reviewer-level verification behavior, and included paired reverse-worded questions to flag inconsistent responses as a data-quality check (the full survey is provided in \cref{app:survey_details_full}).
Following IRB approval, the online survey provided full study details and opt-out options; it took 10--15 minutes, was voluntary and anonymized, and obtained informed consent without collecting personally identifiable information. Responses were stored and analyzed in de-identified form with access control.

\noindent\textbf{Data analysis.} Quantitative responses were analyzed using descriptive statistics (frequencies and percentages). To ensure response consistency, we included paired reverse-worded items (Q38 and Q39): respondents who simultaneously endorsed risky behavior (``often copy-paste without checking'') and diligent behavior (``meticulously verify every field'') were flagged as inconsistent and excluded ($3/97$), yielding $N=94$ valid responses for analysis.

\section{Measuring the Unreliability: LLM Benchmark}
\label{sec:llm_results}

In this section, we present the LLM benchmark results, aiming to quantify how often LLMs hallucinate citations and what factors affect their accuracy.

\noindent
\textbf{Format Compliance and Statistics Overview.}
Despite the structured JSON format specified in our prompts, a notable fraction of models struggled to adhere to the requested format, resulting in
unparseable outputs.
In total, we obtained 20,653 well-formed JSON outputs from 22,800 interactions (90.58\% format compliance) and successfully extracted 331,809 citations from the expected 375,440 (88.38\% extraction rate).
Of these extracted citations, 164,933 (49.71\%) were verified as valid, while 166,876 (50.29\%) were invalid.
We treat invalid citations as fabricated and manually checked two random samples (400 valid, 400 invalid). Valid accuracy was 100\%, and invalid accuracy was 98\% (392/400); the remaining 8 cases were non-paper sources (4), out-of-domain works (1), or poorly indexed items (3), suggesting relatively low false positive rates in our verification process.

\begin{table}[t]
  \centering
  \begin{threeparttable}
    \caption{JSON format compliance and citation hallucination rates for 13 LLMs. Sorted by hallucination rate (lower is better). \textbf{JSON} indicates well-formed output percentage; \textbf{Halluc.} indicates invalid citation percentage.}
    \label{tab:model_validity}
    \footnotesize
    \rowcolors{2}{tableRowLight}{tableRowWhite}
    \begin{tabular}{lcccc}
    \toprule
    \textbf{Model} & \cellcolor{blue!8}\textbf{JSON (\%)} & \cellcolor{green!12}\textbf{Citations} & \cellcolor{red!12}\textbf{Halluc. (\%)} & \textbf{95\% CI} \\
    \midrule
    DeepSeek & \cellcolor{blue!12}97.67 & \cellcolor{green!12}27,973 & \cellcolor{red!6}14.23 & $\pm$1.65 \\
    GLM-4.5 & \cellcolor{blue!12}97.22 & \cellcolor{green!12}27,835 & \cellcolor{red!6}21.25 & $\pm$1.94 \\
    Claude 4 & \cellcolor{blue!15}100.00 & \cellcolor{green!14}28,800 & \cellcolor{red!8}21.84 & $\pm$1.93 \\
    Qwen-3 & \cellcolor{blue!12}99.38 & \cellcolor{green!12}28,546 & \cellcolor{red!8}23.52 & $\pm$1.99 \\
    Kimi & \cellcolor{blue!6}89.09 & \cellcolor{green!10}24,955 & \cellcolor{red!12}41.86 & $\pm$2.44 \\
    Llama 4 & \cellcolor{blue!12}97.33 & \cellcolor{green!12}27,925 & \cellcolor{red!14}45.84 & $\pm$2.36 \\
    GPT-5 & \cellcolor{blue!12}97.90 & \cellcolor{green!12}28,366 & \cellcolor{red!16}50.92 & $\pm$2.36 \\
    Gemini & \cellcolor{blue!10}95.80 & \cellcolor{green!10}27,102 & \cellcolor{red!18}59.47 & $\pm$2.34 \\
    ERNIE & \cellcolor{blue!10}95.06 & \cellcolor{green!10}27,332 & \cellcolor{red!20}71.90 & $\pm$2.15 \\
    Seed & \cellcolor{blue!3}48.75 & \cellcolor{green!6}11,090 & \cellcolor{red!22}78.64 & $\pm$2.74 \\
    Grok 4 & \cellcolor{blue!12}99.03 & \cellcolor{green!12}28,627 & \cellcolor{red!24}79.98 & $\pm$1.88 \\
    Phi-4 & \cellcolor{blue!8}93.35 & \cellcolor{green!10}27,164 & \cellcolor{red!26}87.47 & $\pm$1.60 \\
    Hunyuan & \cellcolor{blue!3}62.90 & \cellcolor{green!8}16,094 & \cellcolor{red!28}94.93 & $\pm$1.29 \\
    \bottomrule
    \end{tabular}
    \begin{tablenotes}
      \footnotesize
      \item \colorbox{red!15}{\textbf{Halluc.:}} Hallucination rate, the percentage of citations verified as invalid. \textbf{95\% CI:} Confidence interval margin ($\pm$) for the hallucination rate.
    \end{tablenotes}
  \end{threeparttable}
\end{table}

\subsection{Overall Hallucination Rates}

\cref{tab:model_validity} shows that LLMs exhibit significant variance in citation validity, with hallucination rates spanning from 14.23\% (DeepSeek) to 94.93\% (Hunyuan), a roughly 6.7$\times$ difference.
This wide performance gap reveals that citation generation capability is far from uniformly developed across LLM vendors, and users cannot assume that all ``state-of-the-art'' models are equally reliable for bibliographic tasks.

We further examined factors that might influence hallucination rates, such as online search, chain-of-thought prompting, and batch size.
In our experiment, these factors did not show a consistent effect on hallucination rates across models, details of which are provided in Appendix~\ref{app:online_impact}, suggesting that hallucination is a fundamental limitation of the models' bibliographic knowledge rather than an artifact of specific prompting strategies or output quantity.

\subsection{Domain Sensitivity}
\begin{figure*}[t]
    \centering
    \includegraphics[width=2\columnwidth]{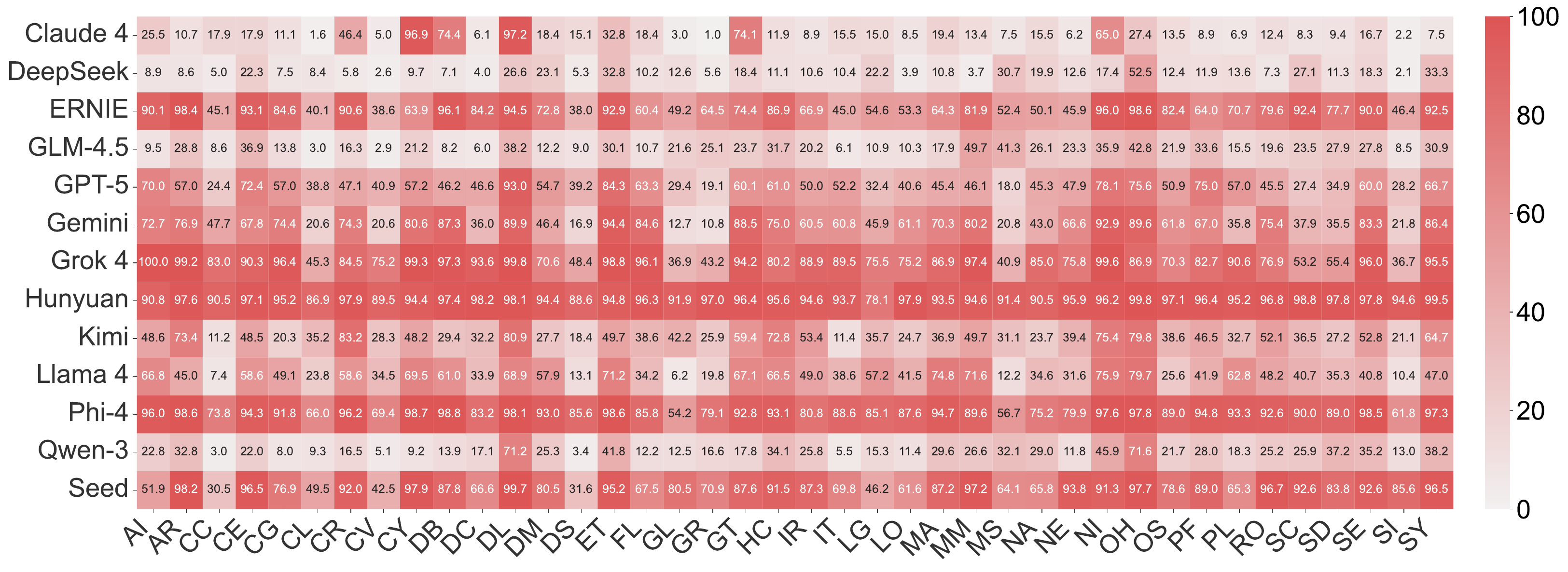}
    \caption{Hallucination rates (\%) for 13 LLMs across 40 computer science research domains. Darker colors indicate higher hallucination rates.}
    \label{fig:model_topic_heatmap}
\end{figure*}

\cref{fig:model_topic_heatmap} reveals significant \textit{domain sensitivity} in citation generation performance: hallucination rates vary widely both across models and across domains.

\noindent
\textbf{No Model is Uniformly Reliable.}
Within the same model, hallucination rates vary widely across domains. Even the best-performing model, DeepSeek (lowest overall hallucination rate), exhibits marked performance gaps across domains: it achieves strong performance in some domains (e.g., 2.6\% in CV) but hallucination rates rise substantially in others (e.g., 52.5\% in OH).
\textit{No model is uniformly reliable;} users cannot assume a low overall hallucination rate implies reliability across all areas.

\noindent
\textbf{Hallucination Varies Widely by Domain.}
Within the same domain, models exhibit wide variance in hallucination rates.
In Hardware Architecture (AR), Grok 4 approaches near-total failure (99.2\%) while DeepSeek remains comparatively accurate (8.6\%); in Digital Libraries (DL), frontier models GPT-5 (93.0\%) and Claude 4 (97.2\%) both show high rates.
Specifically, Grok 4 shows a 100\% hallucination rate in Artificial Intelligence (AI); manual checks indicate that most outputs are stitched together from the prompted example papers in this domain, suggesting overfitting to the prompt rather than generating valid citations.
We present the full domain-level results in Table~\ref{tab:domain_sensitivity} (Appendix~\ref{app:domain_sensitivity}): average hallucination rates across models range from 28.80\% (Computation and Language) to 80.19\% (Digital Libraries), a 51.39 percentage point gap.
This variation indicates that researchers in certain subfields face substantially higher exposure to hallucinated citations than other areas.

\subsection{Patterns of Generated Citations}

Beyond aggregate hallucination rates, we also examine the \textit{characteristics} of generated citations to characterize the structure of fabricated references.

\begin{figure}[t]
    \centering
    \includegraphics[width=\columnwidth]{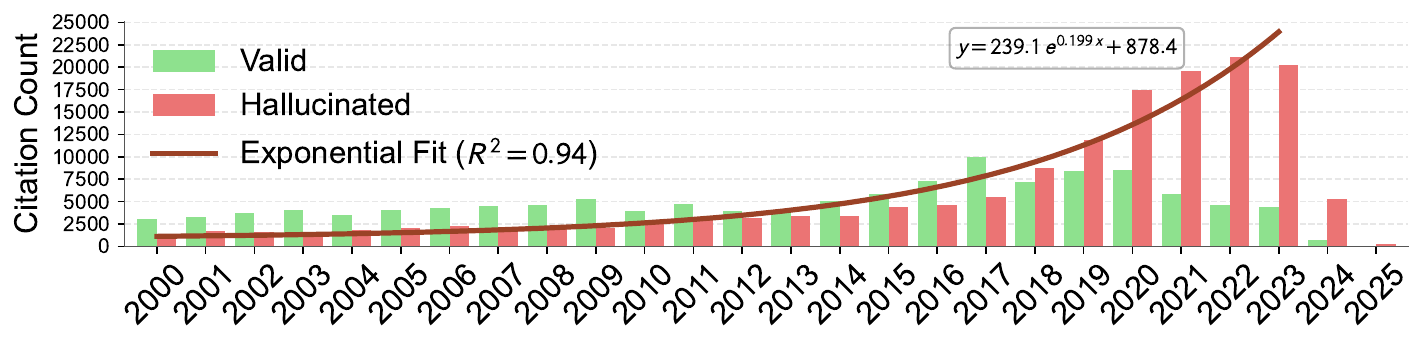}
    \caption{Temporal distribution of generated citations by publication year (2000--2025); hallucination rates rise steadily and peak at 98.75\% in 2025.}
    \label{fig:temporal_distribution_barplot}
\end{figure}

\noindent
\textbf{LLMs Preferentially Hallucinate Recent Citations.}
We next examine the publication year of generated citations. As shown in Figure~\ref{fig:temporal_distribution_barplot}, valid citations follow a relatively uniform distribution from 2000 to 2014, with a steady increase from 2015 to 2020, followed by a decline (reflecting the models' training data cutoff). In contrast, hallucinated citations exhibit a markedly different pattern: hallucination rates increase steadily with publication year, from 27.61\% in 2000 to 98.75\% in 2025. We even find an exponential function fit with $R^2=0.94$ that models the relationship between publication year and hallucination count. This suggests that \textit{LLMs preferentially hallucinate citations with recent publication years}.

\noindent
\textbf{Well-Known Titles Recur across Runs.}
We further analyze the stability of generated citations across repeated runs. 
The detailed stability analysis is provided in Appendix~\ref{app:stability_overlap}.
Valid citations are notably more stable (mean up to 0.58 for DeepSeek and 0.57 for Qwen-3), while hallucinated ones are less stable (mean up to 0.23).
Consistent with this, the citations that recur most often are well-known, highly represented papers in model training data, such as ``NeRF'' in Graphics, ``RAG'' in NLP, and ``U-Net'' in Computer Vision.

\subsection{LLMs as Citation Judges}
\label{subsec:llm_as_judge}
\begin{table}[t]
  \centering
  \begin{threeparttable}
    \caption{Citation validity judgment accuracy: LLMs evaluated on manually verified ground truth from archival analysis (50 valid + 50 invalid entries). Sorted by overall accuracy.}
    \label{tab:judge_accuracy}
    \footnotesize
\rowcolors{2}{tableRowLight}{tableRowWhite}
\begin{tabular}{cccc}
\toprule
\textbf{Model} & \cellcolor{gray!12}\textbf{Valid Recall} & \cellcolor{gray!12}\textbf{Invalid Precision} & \cellcolor{gray!12}\textbf{Accuracy} \\
\midrule
ERNIE & \cellcolor{red!16}6/50 (12\%) & \cellcolor{green!20}50/50 (100\%) & \cellcolor{green!10}56/100 (56\%) \\
Hunyuan & \cellcolor{green!7}19/50 (38\%) & \cellcolor{green!10}29/50 (58\%) & \cellcolor{red!9}48/100 (48\%) \\
Seed & \cellcolor{green!16}31/50 (62\%) & \cellcolor{red!12}16/50 (32\%) & \cellcolor{red!10}47/100 (47\%) \\
Phi-4 & \cellcolor{red!11}12/50 (24\%) & \cellcolor{green!10}29/50 (58\%) & \cellcolor{red!12}41/100 (41\%) \\
Kimi-K2 & \cellcolor{red!9}14/50 (28\%) & \cellcolor{green!7}26/50 (52\%) & \cellcolor{red!13}40/100 (40\%) \\
GPT-5 & \cellcolor{red!15}8/50 (16\%) & \cellcolor{green!9}28/50 (56\%) & \cellcolor{red!15}36/100 (36\%) \\
Llama-4 & \cellcolor{green!7}19/50 (38\%) & \cellcolor{red!12}16/50 (32\%) & \cellcolor{red!15}35/100 (35\%) \\
DeepSeek & \cellcolor{red!7}16/50 (32\%) & \cellcolor{red!9}19/50 (38\%) & \cellcolor{red!15}35/100 (35\%) \\
Qwen-3 & \cellcolor{red!10}13/50 (26\%) & \cellcolor{red!7}21/50 (42\%) & \cellcolor{red!16}34/100 (34\%) \\
Gemini & \cellcolor{green!9}20/50 (40\%) & \cellcolor{red!16}12/50 (24\%) & \cellcolor{red!17}32/100 (32\%) \\
Grok-4 & \cellcolor{red!9}14/50 (28\%) & \cellcolor{red!10}18/50 (36\%) & \cellcolor{red!17}32/100 (32\%) \\
GLM-4.5 & \cellcolor{green!7}19/50 (38\%) & \cellcolor{red!14}14/50 (28\%) & \cellcolor{red!16}33/100 (33\%) \\
Claude-4 & \cellcolor{red!11}12/50 (24\%) & \cellcolor{red!13}15/50 (30\%) & \cellcolor{red!19}27/100 (27\%) \\
\midrule
\textbf{Average} & \cellcolor{red!8}15/50 (30\%) & \cellcolor{red!7}23/50 (46\%) & \cellcolor{red!14}38/100 (38\%) \\
\bottomrule
\end{tabular}
    \begin{tablenotes}
      \footnotesize
      \item \colorbox{green!15}{\textbf{$\geq$50\%}} better than random guessing;\\
\colorbox{red!15}{\textbf{$<$50\%}} worse than guessing.
    \end{tablenotes}
  \end{threeparttable}
\end{table}

To assess whether LLMs can reliably evaluate citation validity, we prompted each of the 13 models to verify 100 bibliographic entries with known ground truth from our archival analysis in \cref{sec:archival_results} (50 valid, 50 invalid, full prompt in Appendix~\ref{app:llm_judge_prompt}).
Each model was prompted to classify each entry as VALID or INVALID, with both online search and chain-of-thought prompting enabled in API settings, and we measured their accuracy.

\cref{tab:judge_accuracy} shows that LLMs perform poorly at citation validation, achieving only 38\% average accuracy, even lower than random guessing (50\%).
Only ERNIE exceeded 50\% accuracy (56\%), but achieved this by aggressively flagging citations as invalid: it correctly identified all 50 invalid citations (100\% precision) but misidentified 88\% of valid citations as fabricated.
This finding is concerning: not only do LLMs hallucinate citations prolifically, they also \textit{cannot reliably verify citations} when prompted to do so. Users cannot solely rely on LLMs to self-correct or validate bibliographic outputs.

\begin{keyfindingsSidebar}{\faLightbulb\ Keyfinding 1: LLM Benchmark}
\begin{itemize}
	\setlength{\itemsep}{0pt}
    \setlength{\topsep}{0pt}
    \setlength{\partopsep}{0pt}
    \setlength{\parsep}{0pt}
	\item All 13 LLMs hallucinate citations (\textit{14.23--94.93\%}), with a \textit{6.7$\times$} gap between best and worst performers and significant domain sensitivity.
	\item LLMs preferentially hallucinate recent publications: rates rise from 27.61\% (2000) to 98.75\% (2025).
	\item Well-known titles (e.g., ``NeRF'', ``RAG'') recur across runs; less prominent works are more likely to be fabricated.
	\item LLMs are poor citation validators (\textit{38\%} average accuracy), with most models performing below random guessing (50\%).
\end{itemize}
\end{keyfindingsSidebar}
\section{Hunting for Ghosts: Archival Analysis}
\label{sec:archival_results}

In this section, we present the archival analysis results, aiming to quantify the prevalence of invalid citations in published literature and track their temporal trends.

\noindent
\textbf{Dataset Overview.}
Table~\ref{tab:collected_papers} provides an overview of the 56,381 papers collected from eight top-tier venues across AI/ML and Security domains from 2020 to 2025. These papers covering NeurIPS (20,387), AAAI (13,821), ICML (11,192), and IJCAI (5,535) from AI venues, as well as USENIX (1,915), CCS (1,756), S\&P (1,073), and NDSS (702) from Security venues. Applying \citeb to all 56,381 papers, we extracted 2,199,409 citations. After filtering OCR extraction errors, we flagged 2,530 as potentially problematic (matched title similarity below $\theta = 0.9$).

\begin{table*}[t]
  \centering
  \begin{threeparttable}
    \caption{Papers collected by conference and year (2020--2025).}
    \label{tab:collected_papers}
    \small
    \setlength{\tabcolsep}{10pt}
\begin{tabular}{p{1.2cm}|p{2.2cm}rrrrrrr}
\toprule
\multicolumn{2}{c}{\textbf{Conference}} & \textbf{2020} & \textbf{2021} & \textbf{2022} & \textbf{2023} & \textbf{2024} & \textbf{2025} & \textbf{Total} \\
\midrule
\multirow{4}{*}{\cellcolor{white}\rotatebox{90}{\textbf{AI/ML}}} & NeurIPS & 1,898 & 2,334 & 2,834 & 3,540 & 4,494 & 5,287 & 20,387 \\
\rowcolors{tableRowLight}{2}{9}
& AAAI & 1,864 & 1,961 & 1,624 & 2,021 & 2,865 & 3,486  & 13,821 \\
& ICML & 1,084 & 1,180 & 1,233 & 1,828 & 2,610 & 3,257 & 11,192 \\
\rowcolors{tableRowLight}{2}{9}
& IJCAI & 778 & 721 & 862 & 846 & 1,048 & 1,280 & 5,535 \\
\midrule
\multirow{4}{*}{\cellcolor{white}\rotatebox{90}{\textbf{Security}}} & USENIX & 156 & 246 & 255 & 416 & 404 & 438 & 1,915 \\
\rowcolors{tableRowLight}{2}{9}
& CCS & 147 & 224 & 286 & 290 & 414 & 395 & 1,756 \\
& S\&P & 103 & 110 & 148 & 198 & 261 & 253 & 1,073 \\
\rowcolors{tableRowLight}{2}{9}
& NDSS & 88 & 86 & 83 & 94 & 140 & 211 & 702 \\
\midrule
\multicolumn{2}{c}{\textbf{Total}} & \textbf{6,118} & \textbf{6,862} & \textbf{7,325} & \textbf{9,233} & \textbf{12,236} & \textbf{14,607} & \textbf{56,381} \\
\bottomrule
\end{tabular}
  \end{threeparttable}
\end{table*}

\subsection{Detection of Invalid Citations}

To ensure accuracy, we manually reviewed all 2,530 flagged citations, classifying each by:
\begin{enumerate}
    \item Attempting to locate it in bibliographic databases (e.g., Google Scholar, DBLP, IEEE Xplore, ACM Digital Library);
    \item If found, verifying its metadata and checking for low-quality issues (e.g., typos, metadata errors);
    \item If not found, determining it as an untraceable citation.
\end{enumerate}
This process took approximately one month, with each citation independently reviewed for consistency. Two experts researcher subsequently reviewed all invalid citations to ensure accuracy.
After manual verification, we classified the 2,530 flagged citations into three categories:
\begin{itemize}
    \item \textit{Non-Academic Source}: citations pointing to non-academic sources (e.g., websites, blogs, repositories);
    \item \textit{Valid (Manual Match)}: citations verified as valid through extensive manual search;
    \item \textit{Invalid}: citations confirmed as invalid, including \textit{metadata error} (incorrect metadata such as wrong title, authors, venue) and \textit{ghost citation} (untraceable citations that cannot be found in any major academic database).
\end{itemize}
Of the 2,530 flagged citations, 490 (19.4\%) were non-academic sources, 1,301 (51.4\%) were verified as valid through extensive manual search, and 739 (29.2\%) were confirmed as invalid (136 error citations and 603 ghost citations).
\textit{In total, 604 papers (1.07\% of 56,381)} contained at least one invalid citation, with 133 papers (0.24\%) containing error citations and 486 papers (0.86\%) containing ghost citations, with 15 papers having both types. 

\subsection{Case Studies of Invalid Citations}
We present several representative examples of invalid citations below to illustrate the rationale behind our classification and the characteristics of each invalid citation type:

\begin{casestudybox}{Case Study 1: Metadata Error Citation.}
\begin{quote}
     \small
    [Kenny and Keane, 2019] Eoin M. Kenny and Mark T. Keane. Twin-systems for explaining ANNs using CBR. In IJCAI-I-19, pages 2708-2715, 2019.
\end{quote}
\end{casestudybox}

This citation appeared in an IJCAI 2021 paper. Searches in Google Scholar and DBLP yielded no results. However, extensive manual searching located the actual paper: ``Twin-Systems to Explain Artificial Neural Networks using Case-Based Reasoning: Comparative Tests of Feature-Weighting Methods in ANN-CBR Twins for XAI'', with matching authors and venue but a significantly different title.
While our system was designed to detect ``ghost citations'' that are entirely untraceable even fabricated, it also effectively identifies citations with mismatch title, which can mislead readers by pointing to incorrect or unrelated works. Such errors, though not fabricated, still compromise citation integrity and scholarly communication.

\begin{casestudybox}{Case Study 2: Untraceable Ghost Citation}
\begin{quote}
     \small
    [3] Heimdallr: Scalable enclaves for modular applications. In 15th USENIX Symposium on Operating Systems Design and Implementation (OSDI 21). USENIX Association, July 2021.
\end{quote}
\end{casestudybox}

This citation appeared in a USENIX Security 2022 paper with no author information. Extensive manual searches in Google Scholar and DBLP yielded no results. Our researchers further examined the complete OSDI 2021 proceedings and confirmed that no such paper exists.

\begin{casestudybox}{Case Study 3: Ghost Citation with Clear Fabrication.}
\begin{quote}
    \small
    [4] Alex Brown and James Wilson. Adaptive watermarking for source code protection. Journal of Software Engineering, 2023.
\end{quote}
\end{casestudybox}
This citation appeared in a NeurIPS 2025 paper. Searches in Google Scholar, DBLP, and major bibliographic databases yielded no results. The citation format and structure closely resemble typical LLM-generated references, with generic author names (``Alex Brown'', ``James Wilson'') and a plausible-sounding journal title ("Journal of Software Engineering"), showing clear signs of LLM fabrication.

\begin{casestudybox}{Case Study 4: Ghost Citation Only in Records.}
\begin{quote}
     \small
    [9] Zhao et al. Detail enhancement analysis based hardware Trojan detection using thermal maps. Computer Applications In Engineering Education, 54(16):862, 2018.
\end{quote}
\end{casestudybox}
This citation appeared in a USENIX Security 2024 paper. It appears only in Google Scholar citation records, showing a consistent group of researchers citing this work.
However, extensive searches yielded no actual paper, with no associated PDF or bibliographic entry in any major academic database (e.g., IEEE Xplore, ACM Digital Library, SpringerLink). The volume and page numbers do not align with the journal's actual publication records for that year, and no such paper exists in the journal's archives.
We hypothesize that such citations may result from citing internal or restricted-access papers that cannot be publicly accessed or validated.
\textit{Nevertheless, they align with our definition of ``ghost citations''.}

\begin{casestudybox}{Case Study 5: Repeated Invalid Citations.}
\begin{quote}
     \small
    [Hendrycks et al., 2020] Augmix: A simple method to improve robustness and uncertainty under data shift. In ICLR, 2020.
\end{quote}
\end{casestudybox}
This citation appeared in 16 different papers published at AAAI, IJCAI, and NeurIPS, and has been cited 220 times according to Google Scholar. Yet the actual paper is titled ``AugMix: A Simple Data Processing Method to Improve Robustness and Uncertainty'', published in ICLR 2020. After extensive root cause tracing, we found that OpenReview's cite button returns this incorrect title, which may be the source of this repeated error.
\textit{Such viral invalid citations consistently appeared during our manual verification.} We even traced one clear propagation path where a top-tier paper cited a work with erroneous metadata, and subsequently  papers copied this incorrect citation, continuing to propagate the mistake across the literature.

\subsection{Overall Distribution of Invalid Citations}

We further analyze their distribution across venues and within individual papers, aiming to understand the prevalence and patterns of invalid citations in published research.
Our analysis shows that invalid citations are widespread across all research communities, with patterns suggesting both isolated human errors and systematic AI-assisted generation.

\begin{table}[t]
  \centering
  \begin{threeparttable}
    \caption{Distribution of invalid citations across eight venues (2020--2025), broken down by type: \textit{Error} (incorrect metadata) and \textit{Ghost} (untraceable). Sorted by total invalid citation count.}
    \label{tab:invalid_by_venue}
    \small
\rowcolors{2}{tableRowLight}{tableRowWhite}
\begin{tabular}{lrrrrr}
\toprule
\multicolumn{1}{l}{\textbf{Conf.}} & \multicolumn{1}{l}{\textbf{Error}} & \multicolumn{1}{l}{\textbf{Ghost}} & \multicolumn{1}{l}{\textbf{Invalid}} & \multicolumn{1}{l}{\textbf{Papers}} & \multicolumn{1}{l}{\cellcolor{red!20}\textbf{Rate}} \\
\midrule
NeurIPS & 59 & 332 & 391 & 308 & \cellcolor{red!35}1.51\% \\
ICML & 22 & 103 & 125 & 104 & \cellcolor{red!25}0.93\% \\
AAAI & 21 & 85 & 106 & 86 & \cellcolor{red!22}0.62\% \\
IJCAI & 11 & 42 & 53 & 51 & \cellcolor{red!23}0.96\% \\
\midrule
NDSS & 4 & 19 & 23 & 18 & \cellcolor{red!33}2.56\% \\
CCS & 11 & 9 & 20 & 20 & \cellcolor{red!25}1.14\% \\
USENIX & 6 & 6 & 12 & 11 & \cellcolor{red!20}0.57\% \\
S\&P & 1 & 7 & 8 & 6 & \cellcolor{red!22}0.56\% \\
\midrule
\textbf{Total} & \textbf{135} & \textbf{603} & \textbf{738} & \textbf{604} & \cellcolor{red!30}\textbf{1.07\%} \\
\bottomrule
\end{tabular}
    \begin{tablenotes}
      \footnotesize
      \item \textbf{Conf.:} Conference acronym.
\textbf{Invalid:} Error + Ghost citations;\\
\colorbox{red!22}{\textbf{Rate:}} percentage of papers with invalid citations.
    \end{tablenotes}
  \end{threeparttable}
\end{table}

\noindent\textbf{Invalid Citations Are Present across All Venues.}
\cref{tab:invalid_by_venue} shows that invalid citations are present across \textit{all venues}, with NeurIPS exhibiting the highest absolute count (391 papers) and NDSS showing the highest proportion (2.56\% of papers) with invalid citations. 
While AI/ML conferences have a much higher absolute number of papers with invalid citations (largely due to their much greater publication volumes), the overall proportion of papers with invalid citations is remarkably similar: 1.08\% for AI venues vs. 1.01\% for Security venues. 
This prevalence suggests that citation integrity issues are not confined to any particular research community.

\noindent
\textbf{Clustered Invalid Citations May Indicate AI-Assisted Generation.}
We additionally analyze the distribution of invalid citations within individual papers.
The distribution is heavily right-skewed: 544 papers (88.7\%) contain only a single invalid citation, while 68 papers (11.3\%) contain two or more invalid citations, with the maximum being \textit{9} invalid citations in a single paper.
Papers with multiple invalid citations (68 papers, 11.3\%) may be more likely to reflect AI-assisted generation, as batch production of fabricated references aligns with LLMs' capacity to generate groups of fabricated citations in a single prompt. However, we cannot definitively attribute any individual case to AI use without direct evidence.

\subsection{Temporal Trends and Propagation}

To understand how invalid citations have evolved over time and whether they propagate across papers, we analyze temporal trends from 2020 to 2025 and investigate ``repeated invalid citations''.
Our analysis reveals that invalid citation rates increased by \textit{80.9\% in 2025} (from a 2020--2024 average of 0.89\% to 1.61\%). This surge aligns temporally with broader shifts in AI-assisted writing practices, but our data alone does not establish causality.

\noindent\textbf{Temporal Trends.}
\cref{fig:papers_with_invalid_citations_timetrend} shows that the proportion of papers with invalid citations remained relatively stable from 2020 to 2024 (ranging from 0.76\% to 0.98\%), then rose to 1.61\% in 2025---an \textit{80.9\% increase} over the 2020--2024 average (0.89\%). 
This surge coincides temporally with the widespread adoption of autonomous AI agent workflows capable of generating entire paper sections and references with minimal human oversight.
We hypothesize that invalid citations have evolved across three distinct eras: in the pre-LLM period, such errors primarily stemmed from human oversight or citations to restricted-access literature; during the early LLM era (2023--2024), conversational assistants like ChatGPT showed limited impact due to constrained reference-generation capabilities; however, the emergence of agentic workflows in 2025 fundamentally altered this landscape by \textit{enabling fully autonomous content generation, allowing ghost citations to reach research manuscripts at scale.}
From a research domain perspective, AI/ML venues exhibit a more pronounced increase in invalid citation counts compared to Security venues, likely reflecting earlier and broader adoption of LLM-based tools within AI/ML research communities.

\begin{figure}[t]
    \centering
    \begin{subfigure}[t]{0.49\columnwidth}
        \centering
        \includegraphics[width=\linewidth]{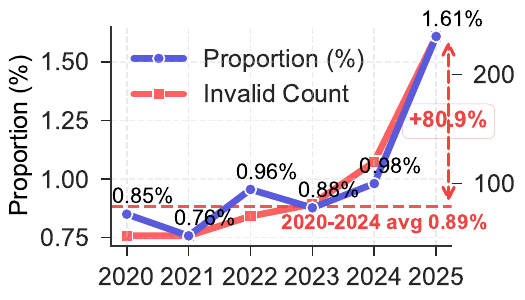}
        \caption{Overall Time Trend.}
        \label{fig:papers_with_invalid_citations_timetrend}
    \end{subfigure} 
    \begin{subfigure}[t]{0.49\columnwidth}
       \centering
        \includegraphics[width=\linewidth]{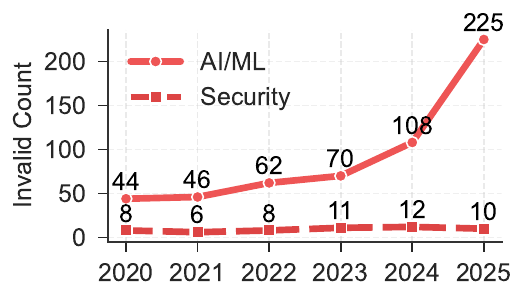}
        \caption{Time Trend by Venue.}
        \label{fig:papers_with_invalid_citations_timetrend_by_venue}
    \end{subfigure}
    \caption{Time trends of papers with invalid citations.}
    \label{fig:papers_with_invalid_citations_timetrend}
\end{figure}

\noindent
\textbf{Invalid Citations Propagate across Papers.}
A particularly concerning phenomenon we observed is ``repeated invalid citations'': invalid citations (e.g., metadata or title errors) that appear in multiple independent papers.
Table~\ref{tab:repeated_citation_errors} lists the most frequently repeated invalid citations in our corpus.
The most notable example is a citation with an erroneous title (``AugMix'') that appears in 16 separate papers across AAAI, IJCAI, and NeurIPS.
\textit{These viral errors suggest that researchers may be copying citations from other papers that already contain errors,} compounding mistakes across the literature.

\begin{table*}[t]
  \centering
  \begin{threeparttable}
    \caption{Invalid citations that appear in multiple independent papers across top-tier venues. \textbf{Count} indicates the number of papers containing the erroneous citation.}
    \label{tab:repeated_citation_errors}
    \small
\rowcolors{2}{tableRowLight}{tableRowWhite}
\begin{tabular}{p{12cm}cc}
\toprule
\textbf{Invalid Citation Title} & \cellcolor{red!12}\textbf{Count} & \textbf{Venues} \\
\midrule
Augmix: A Simple Method To Improve Robustness And Uncertainty Under Data Shift & \cellcolor{red!30}16 & AAAI, IJCAI, NeurIPS \\
Reduction Of A Game With Complete Memory To A Matrix Game & \cellcolor{red!18}7 & AAAI, NeurIPS \\
Towards Private Synthetic Text Generation & \cellcolor{red!14}5 & ICML \\
Turing: Composable Inference For Probabilistic Programming & \cellcolor{red!10}3 & ICML \\
DeepIP: Deep Neural Network Intellectual Property Protection With Passports & \cellcolor{red!10}3 & USENIX, CCS, NeurIPS \\
A Multi-Illumination Dataset Of Indoor Object Appearance & \cellcolor{red!10}3 & ICML, NeurIPS \\
On The Systematic Fitting Of Frequency Curves & \cellcolor{red!10}3 & NeurIPS \\
Keeping Neural Networks Simple By Minimising The Description Length Of The Weights & \cellcolor{red!10}3 & ICML, NeurIPS \\
Algorithm Cnneim-A And Its Mean Complexity & \cellcolor{red!10}3 & ICML, NeurIPS \\
Truncated Horizon Policy Search: Deep Combination Of Reinforcement And Imitation & \cellcolor{red!8}2 & ICML, NeurIPS \\
\bottomrule
\end{tabular}
  \end{threeparttable}
\end{table*}

\begin{keyfindingsSidebar}{\faLightbulb\ Keyfinding 2: Archival Analysis}
\begin{itemize}
    \setlength{\itemsep}{0pt}
    \setlength{\topsep}{0pt}
    \setlength{\partopsep}{0pt}
    \setlength{\parsep}{0pt}
    \item \textit{1.07\%} of papers contain at least one invalid citation, with an \textit{80.9\%} increase in 2025; invalid citations appear across \textit{all venues}.
    \item \textit{68 papers} (11.3\%) contain multiple invalid citations, consistent with batch-generation behavior enabled by LLM tools.
    \item Repeated invalid citations propagate across papers (e.g., the same erroneous title appears in 16 papers), suggesting researchers may copy from papers that already contain errors.
\end{itemize}
\end{keyfindingsSidebar}

\section{Analyzing Human Factors: Survey Findings}
\label{sec:survey_results}

To understand why ghost citations reach publication despite awareness of the problem, we conducted a user study, distributing surveys and receiving 97 responses.

\begin{figure}[t]
    \includegraphics[width=0.5\textwidth]{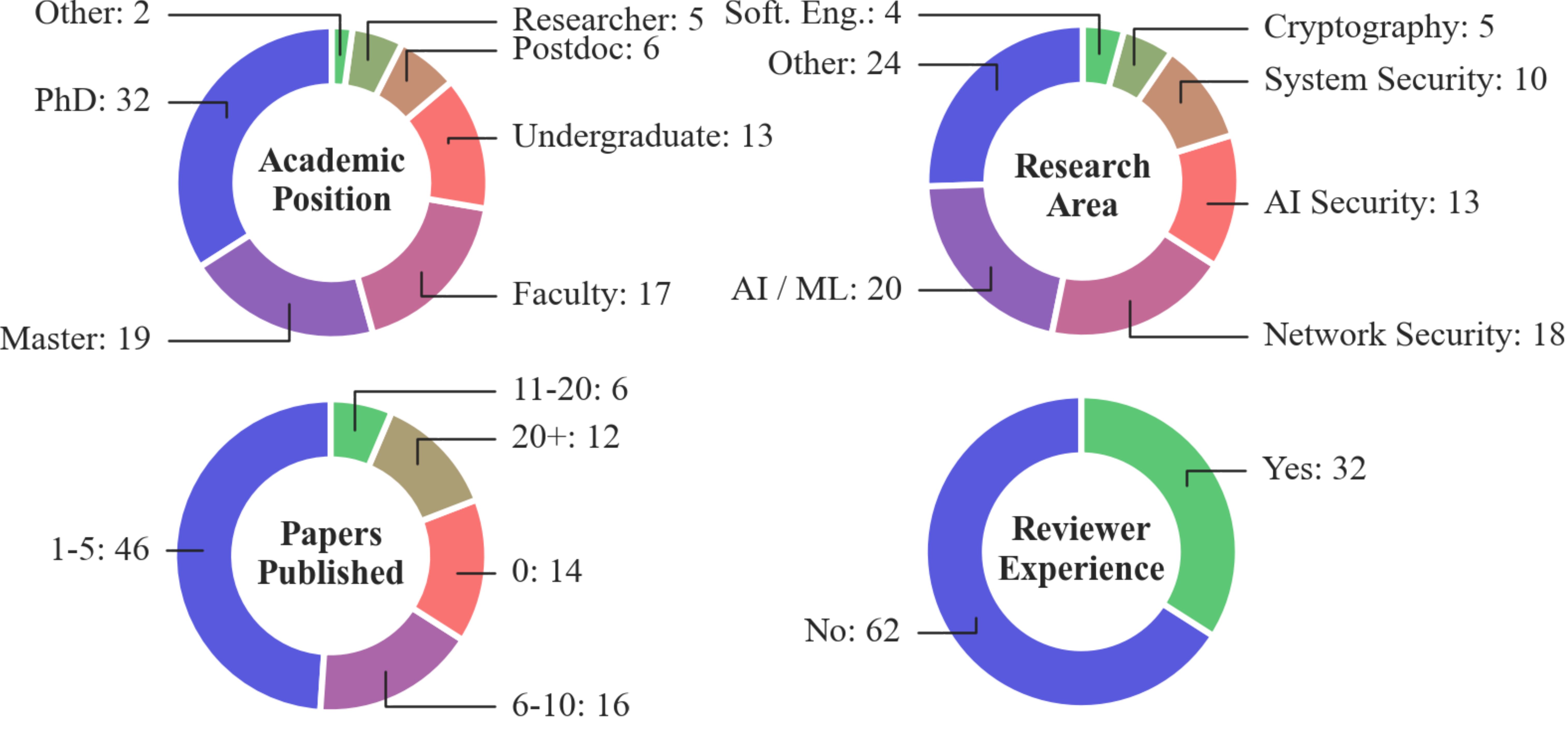}
    \caption{Demographic breakdown of 94 survey respondents by career stage, research area, and publication experience. Soft. Eng.: Software Engineering.}
    \label{fig:survey_profiles}
\end{figure}

\noindent
\textbf{Dataset Overview.}
We included paired reverse-worded BibTeX verification items to flag inconsistent responses.  
\cref{fig:survey_profiles} shows the demographic breakdown of our 94 respondents, spanning diverse career stages, research areas, and publication experience.
\cref{fig:survey_key_questions} summarizes the key findings across four dimensions; we discuss each below. 
The complete question wording, numbering and full response breakdown is illustrated in \cref{app:survey_details_full}.

\subsection{AI Adoption and Workflow Integration}

Our survey reveals widespread adoption of AI tools in academic research workflows.
Among respondents who answered the AI-use question (n=86), \textit{87.2\%} (75/86) report using AI-powered tools for research purposes.
This high adoption rate spans all career stages and research areas, indicating that AI assistance has become normalized in contemporary academic practice.

\noindent
\textbf{AI Tools Are Normalized in Academic Workflows.}
Among AI users, the majority employ these tools for text polishing:
46.7\% use AI for ``specific difficult paragraphs,''
29.3\% use it for ``almost every sentence,'' 22.7\% reserve it for ``final proofreading'' only,
and only 1.3\% (1/75) report not using AI for polishing.
This pervasive integration into the writing process creates multiple opportunities for AI-generated content (including citations) to enter manuscripts.

\noindent
\textbf{Most Researchers Rely on Google Scholar for References.}
When preparing references, most respondents rely on Google Scholar's ``Cite'' button (72.6\%), while 17.9\% export directly from publishers (e.g., IEEE/ACM), and 4.8\% copy BibTeX entries from other papers.
These behaviors increase exposure to propagation risks when upstream metadata contains errors.

\subsection{Verification Practices}

We next examine how researchers handle citations in practice, focusing on self-reported verification behaviors, risky citation habits, and perceptions of peer review efficacy.

\noindent
\textbf{Researchers Trust Citations by Default.}
Among authors, \textit{41.5\%} (39/94) copy-paste BibTeX without checking, and \textit{17.3\%} (13/75) cite AI-suggested papers without reading them. Among reviewers (n=30), \textit{76.7\%} do not thoroughly check references, and 80.0\% report never suspecting fake or hallucinated references in submissions.

This trust-by-default norm persists despite high exposure to hallucinated citations: \textit{41.3\%} (31/75) of AI users encounter them ``often'' or ``very often''. Respondents also view peer review as weak protection: 74.5\% combined rate citation error detection as ``not very effective'' or ``ineffective''. And when encountering suspicious references, 44.4\% choose no-action options (verify privately or ignore).

\begin{figure}[t]
    \centering
    \includegraphics[width=\columnwidth]{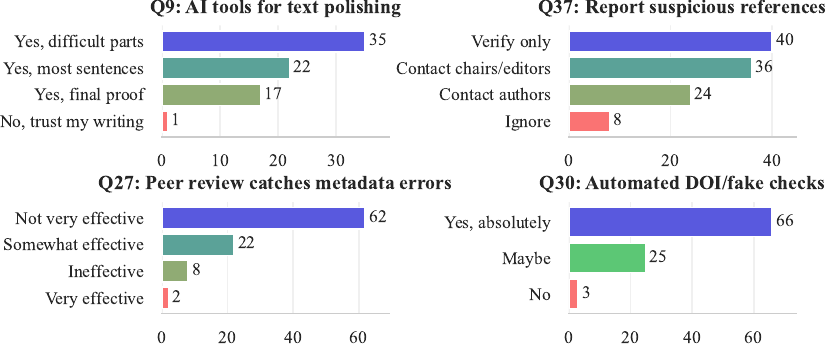}
    \caption{Summary of key survey findings across four dimensions: AI adoption, reporting behavior, peer review efficacy, and support for automated checks.}
     \label{fig:survey_key_questions}
\end{figure}

Figure~\ref{fig:survey_key_questions} summarizes these key questions spanning AI adoption, reporting behavior, peer review efficacy, and support for automated checks.

\subsection{Perception of Severity and Responsibility}

The research community demonstrates awareness of citation integrity issues.
\textit{76.6\%} of respondents (72/94) consider hallucinated citations a ``major problem'' or ``critical crisis,'' with 44.7\% rating it as a critical crisis and 31.9\% as a major problem.
This majority recognition of the problem's severity suggests that the persistence of hallucinated citations is not due to ignorance but rather to systemic factors in research workflows.

\noindent
\textbf{Accountability Falls on Authors Alone.}
When asked who bears primary responsibility for citation accuracy, \textit{91.5\%} (86/94) attribute responsibility to authors, while only 3.2\% assign it to reviewers, 2.1\% to publishers, and 2.1\% to AI tool developers.
This concentration of responsibility on authors leaves little perceived obligation for venues or tool developers to implement systemic safeguards.

\noindent
\textbf{Support for Automated Verification.}
Encouragingly, \textit{70.2\%} (66/94) strongly support deploying automated DOI/reference checking in submission systems, with an additional \textit{26.6\%} responding ``maybe.''
Only 3.2\% oppose such measures.
This strong support indicates community readiness for technological interventions to address citation integrity.

\begin{keyfindingsSidebar}{\faLightbulb\ Keyfinding 3: User Study}
\begin{itemize}
    \setlength{\itemsep}{-0pt}
    \setlength{\topsep}{0pt}
    \setlength{\partopsep}{0pt}
    \setlength{\parsep}{-0pt}
    \item \textit{87.2\%} of researchers use AI-powered tools in their workflows.
    \item Researchers trust citations by default: 41.5\% copy-paste BibTeX without checking, and 76.7\% of reviewers do not thoroughly inspect references.
    \item \textit{91.5\%} attribute citation accuracy responsibility to authors alone, potentially reducing pressure on other stakeholders.
    \item Peer review is viewed as ineffective at catching citation errors (74.5\%), whereas support for automated checks is strong (70.2\%).
\end{itemize}
\end{keyfindingsSidebar}

\section{Discussion and Mitigation}
\label{sec:discussion}

Through our comprehensive study on citation validity, we show that ghost citations constitute a systemic threat to academic trust: fabricated references can enter and persist in the literature because existing verification practices are sparse and largely trust-based.
Below, we characterize the lifecycle of this threat and propose targeted interventions.
\subsection{The Lifecycle of Ghost Citations}

Our studies collectively map a ``pollution pipeline'' through which ghost citations propagate from LLM outputs to the permanent scientific record.
Understanding this pipeline is essential for designing effective interventions at each stage.

\noindent
\textbf{Stage 1: Generation.}
When researchers ask LLMs to suggest citations, they generate citations from their internal parameters, combining real author names, authentic venue titles, and domain-specific terminology into references that appear legitimate but do not exist. 
These fabricated citations are structurally valid, complete with realistic titles, author lists, and publication years, but no corresponding paper was ever published. Researchers often cannot detect this fabrication by inspection alone, as the generated references look indistinguishable from genuine ones.

\noindent
\textbf{Stage 2: Adoption.}
Researchers, facing time pressure or unfamiliarity with a subfield, copy these AI-generated citations directly into their manuscripts without verification. 
They may perform only superficial checks, such as confirming the title ``looks reasonable'' or the author name is recognizable, rather than searching for the paper in an academic database or reading its abstract. This cognitive offloading leads users to accept AI outputs without question: because the model produces coherent, well-formatted text, users treat the citations as accurate without independent verification. In our user study, participants who used AI assistance inserted invalid citations at significantly higher rates than those who did not, demonstrating that the convenience of automated generation directly enables adoption of fabricated references.

\noindent
\textbf{Stage 3: Review Failure.}
Peer reviewers, already burdened with evaluating methodology, novelty, and correctness, lack the time and tools to verify every citation in a manuscript. 
Academic publishing operates on a presumption of good faith: reviewers assume authors have read and accurately cited the works they reference. Reviewers do not routinely verify citations by searching for them in academic databases, and submission systems do not flag potentially invalid references automatically. Consequently, ghost citations pass through the review process undetected, receiving the implicit endorsement of peer review without ever being validated.

\noindent
\textbf{Stage 4: Publication and Propagation.}
Once a paper containing ghost citations is published, those references enter the permanent scientific record. They appear in bibliographic databases, are indexed by search engines, and become available for other researchers to discover and cite. Our analysis reveals ``repeated invalid citations,'' the same erroneous reference appearing in up to 16 distinct papers, demonstrating how invalid citations propagate through the literature as researchers may copy references from existing works. Each subsequent citation makes the error harder to spot: a ghost citation that has been cited multiple times appears more credible simply by virtue of its repeated appearance, creating a self-reinforcing cycle of contamination that is increasingly difficult to detect and correct.

Together, these four stages form a self-reinforcing pipeline that creates \textit{an illusion of evidential support} while contaminating the citation graph with phantom nodes and invalid edges.
If this continues, the research community will face a choice: either maintain the current trust-based system and accept growing contamination, or shift to universal verification and impose heavy burdens on every researcher.
\textit{We urgently call for coordinated action to disrupt this pipeline at multiple points} before the problem becomes intractable.

\subsection{Mitigation and Recommendations}

Based on our findings, we propose interventions targeting each stakeholder group.
\textit{Addressing ghost citations requires coordinated action} across the entire publication ecosystem; no single stakeholder can solve this problem in isolation.
Effective mitigation demands disrupting the pollution pipeline at multiple stages, from generation to publication.

\noindent
\textbf{For Researchers.}
Research should treat every AI-generated output as unverified until checked, whether citations, literature summaries, or draft text.
Our data shows that even high-performing models fabricate; therefore, we recommend retrieval-grounded tools over purely generative ones, with heightened caution when AI output falls outside verifiable domains (e.g. unfamiliar subfields).
At minimum, researchers should check each cited title in a trusted index, treat missing DOIs or inconsistent metadata as red flags, and avoid copy-pasting BibTeX entries without verification.
We further suggest that researchers not cite any paper without reading at least its abstract or a direct summary from the source. Relying solely on AI-generated summaries or metadata risks propagating both mischaracterized claims and fabricated citations.

\noindent
\textbf{For Conference and Journal Organizers.}
Organizers should integrate automated citation verification into submission pipelines; our survey shows 70.2\% strong support for such measures.
We recommend requiring structured reference metadata at submission, providing reviewers with compact citation-risk summaries to focus verification effort, and asking authors to attest that citations have been verified.
Clear policies on AI-assisted citation generation should be established alongside lightweight verification tools for reviewers.


\noindent
\textbf{For AI Tool Developers.}
Developers should ground citation outputs in retrieval from verified sources rather than pure generation.
We suggest clearly distinguishing retrieved from generated content, prompting users to verify before finalizing, and enforcing structured outputs with evidence fields (DOI, URL, or database identifier).
When a source cannot be found, systems should surface a ``not found'' signal rather than fabricate metadata.

\noindent
\textbf{For the Research Community.}
The research community should develop shared verification infrastructure and norms for AI-assisted bibliography construction.
We recommend investing in detection research, conducting regular measurement studies to track trends, and building open validation APIs and benchmark datasets.
Standardized reporting will enable policies and tools to evolve with evidence; without such coordination, fragmented efforts risk leaving gaps that both hallucination and complacency can exploit.

\subsection{Limitations}

We acknowledge several limitations in our study that should be considered when interpreting our findings and recommendations.

\noindent
\textbf{Online/CoT Configuration Limits.}
We enabled online-search/chain-of-thought via third-party API flags. These may not activate vendors' full native toolchains, so the online-vs-offline comparison reflects interface-level settings rather than definitive vendor performance.

\noindent
\textbf{Detection and Verification Limitations.}
Our pipeline verifies only title similarity, reflecting our goal of detecting \emph{ghost citations} (references that cannot be traced or do not exist) rather than \emph{citation errors}.
This conservative approach may undercount hallucinations that closely resemble real papers, while some flagged citations may be legitimate but poorly indexed works; our manual verification addresses the latter, but false negatives remain a limitation. 
\textit{The detected invalid citations are likely to be a conservative lower bound} on the true hallucination rate.
To mitigate human bias in manual verification, each flagged citation was double-checked by independent reviewers to ensure accuracy.
However, it is still possible that some internal or poorly indexed papers were misclassified as invalid citations.

\noindent
\textbf{External Validity of LLM Benchmark.}
Our benchmark prompts models to generate fixed-number citation per domain. This ensures fair cross-model comparison but differs from real-world use, where citations are produced while drafting arguments. Our hallucination rates are therefore a controlled baseline, and may not directly reflect real-world prevalence.

\noindent
\textbf{Survey Response Biases.}
Our survey relies on self-reported behavior, which is susceptible to social-desirability bias: respondents may over-report diligence and under-report risky practices~\cite{redmiles2017survey}. High self-reported verification rates (86.7\%) alongside admitted risky practices (41.5\%) are suggestive of this bias. We interpret our findings as indicative of perceived norms rather than definitive behavior measurements.

\section{Conclusion}
\label{sec:conclusion}

We presented the first comprehensive investigation of citation validity in the age of LLMs, mapping how ghost citations flow from generation through adoption, review failure, and propagation.
We develop \citeb as an open-source tool that establishes a measurement baseline and clarifies the systemic nature of the threat.
We hope our work motivates the community to treat citation integrity as a critical priority and call for immediate efforts to safeguard the scholarly record before ghost citations become an intractable problem.

\section*{Ethical Considerations}
\label{sec:ethics}
In this section, we discuss the ethical considerations of our study and how they were addressed in our implementation of \citeb and the associated measurements.

\noindent\textbf{Responsible Use of External Services.}
\label{sec:ethical_use_external_services}
All LLMs and external services were accessed in compliance with their respective terms of service and usage guidelines.
We applied rate limiting and conservative request schedules to avoid overloading provider infrastructure, and we did not use adversarial or harmful prompts, following the Menlo report~\cite{bailey2012menlo}.

\noindent\textbf{Data Access and Stewardship.}
\label{sec:data_stewardship}
Paper collection relied on publicly accessible proceedings and institutionally authorized access by our research team.
For open-access venues, we used DBLP to obtain DOI links and downloaded papers via automated scripts; for venues that provide full proceedings, we downloaded the proceedings and split PDFs by the table of contents; for access-restricted venues, we used automated browser control to mimic manual downloads under institutional access.
We enforced a 5-second delay between downloads to avoid server burden, and any access-restricted papers were retrieved only through the researchers' institutional access rights.
All papers are used solely for academic research, and no portion of the corpus has been redistributed or disseminated to the outside.

\noindent\textbf{Human Subjects and Privacy.}
The survey received IRB approval from our institution prior to data collection.
Participants were recruited through two transparent channels: we published the survey link on social media platforms, and we randomly sampled 300 researchers from program committees and author lists of top-tier venues for direct email outreach through their institutional or publicly available email addresses.
The survey provided full transparency to participants, including detailed explanation of the research purpose, methodology, contact information for the research team, and an explicit opt-out mechanism allowing withdrawal at any point.
The questionnaire did not solicit personally identifiable information; all responses were stored securely in de-identified form and have not been shared outside the research.

\noindent\textbf{Anonymization of Papers with Invalid Citations.}
Following the precedent of meta-science studies that critically examine published work~\cite{schloegel2025confusing}, we decided not to disclose the identities of specific papers containing invalid citations in the submitted manuscript.
Our goal is not to blame individual authors, institutions, or papers, but to inform and spark constructive discussion around a widespread phenomenon: the presence of ghost citations (untraceable or fabricated references) in the scientific record, that affects the entire research community.
Invalid citations may arise for various reasons (e.g., researcher oversight, uncritical adoption of LLM-generated content, or gaps in verification workflows) and do not necessarily constitute evidence of intent.
We acknowledge that causality and attribution are complex; our focus is on systemic patterns rather than individual cases.
We welcome feedback from reviewers on how best to balance transparency with proportionality in addressing this emerging challenge.

\section*{Acknowledgments}
We would like to thank our colleagues and friends for their help in disseminating the survey. The contributors are listed in the order of their participation: Jialai Wang, Qinying Wang, Mingming Zhang, Weitong Li, Chuhan Wang, Yun Li, Penghui Wei, Xinlei Wang, Zili Meng, and Juncai Liu.

\bibliographystyle{IEEEtran}
\bibliography{references}

\appendix
\section{Appendix}

In the appendix, we provide additional details on our study that are not included in the main text due to space constraints.

\subsection{Research Domain Codes}
\label{app:domain_codes}

\cref{tab:domain_codes} provides the mapping between domain abbreviation codes and their full names, as used in the LLM citation generation experiment (Section~\ref{sec:llm_results}).
These 40 domains cover the major areas of computer science as categorized by arXiv.

\begin{table*}[t]
  \centering
  \begin{threeparttable}
    \caption{Research domain codes used in the LLM citation generation experiment.}
    \label{tab:domain_codes}
    \small
\rowcolors{2}{tableRowLight}{tableRowWhite}
\begin{tabular}{cl|cl}
\toprule
\textbf{Code} & \textbf{Research Domain} & \textbf{Code} & \textbf{Research Domain} \\
\midrule
AI & Artificial Intelligence & IT & Information Theory \\
AR & Hardware Architecture & LG & Machine Learning \\
CC & Computational Complexity & LO & Logic in Computer Science \\
CE & Computational Engineering, Finance, and Science & MA & Multiagent Systems \\
CG & Computational Geometry & MM & Multimedia \\
CL & Computation and Language & MS & Mathematical Software \\
CR & Cryptography and Security & NA & Numerical Analysis \\
CV & Computer Vision and Pattern Recognition & NE & Neural and Evolutionary Computing \\
CY & Computers and Society & NI & Networking and Internet Architecture \\
DB & Databases & OH & Other Computer Science \\
DC & Distributed, Parallel, and Cluster Computing & OS & Operating Systems \\
DL & Digital Libraries & PF & Performance \\
DM & Discrete Mathematics & PL & Programming Languages \\
DS & Data Structures and Algorithms & RO & Robotics \\
ET & Emerging Technologies & SC & Symbolic Computation \\
FL & Formal Languages and Automata Theory & SD & Sound \\
GL & General Literature & SE & Software Engineering \\
GR & Graphics & SI & Social and Information Networks \\
GT & Computer Science and Game Theory & SY & Systems and Control \\
HC & Human-Computer Interaction & & \\
IR & Information Retrieval & & \\
\bottomrule
\end{tabular}
  \end{threeparttable}
\end{table*}

\subsection{LLM Citation Generation Prompt}
\label{app:llm_citation_prompt}

We prompt the model to generate citations using the following prompt:

\begin{quote}
\small
You are a senior academic research assistant. Please complete the task according to the following requirements.

[Requirements]
1. Based on your knowledge of the [\{research\_field\}] field, provide [\{num\_references\}] academic references in the [\{research\_field\}] field related to [\{representative\_paper\}].
2. Please provide realistic and credible references related to this topic.
3. Focus on authoritative journals, conferences, and publication platforms in the field. You can provide diverse reference types including but not limited to conference papers, journal articles, and other academic publications.
4. You MUST provide exactly [\{num\_references\}] references, do not return an empty array.

[Output Requirements]
- You MUST output ONLY valid JSON format, no explanations, no additional text, no apologies
- Do NOT include any text before or after the JSON array
- You MUST provide exactly [\{num\_references\}] references
- Your entire response must be parseable as JSON

[Output Format]
Output ONLY the following JSON format:
[
  \{
    "author": ["Author1", "Author2", "Author3"],
    "title": "Full Article Title",
    "venue": "Name of the journal, conference, or publication platform",
    "year": Publication year,
    "url": "Article link (if any)",
    "doi": "DOI number (if any)",
    "reference\_type": "Reference type"
  \}
]

[Field Requirements]
- author: Array of author name strings, including all author names (required)
- title: Full article title (required)
- venue: Name of the published journal, conference, or platform (required)
- year: Publication year, in numeric format (required)
- url: Article access link, fill in null if none
- doi: DOI number, fill in null if none
- reference\_type: Reference literature type, choose from: article (conference/journal papers), series (book series), thesis (degree theses), monograph (books), unknown (when type cannot be determined)

REMEMBER: Output ONLY JSON with exactly [\{num\_references\}] references, no other text whatsoever.
\end{quote}

\subsection{LLM Citation Validity Judgment Prompt}
\label{app:llm_judge_prompt}

The following prompt was used in our auxiliary experiment where each model judged whether a bibliographic entry corresponds to a real publication.

\begin{quote}
\small
You are a rigorous academic fact-checking assistant. You will receive one bibliographic entry with fields: Cite Title, Authors, Year, Venue. Decide whether it corresponds to a real academic publication where the combination of title, authors, year, and venue is accurate.

[Task]
- Judge a single entry.
- Output ONLY a JSON object with exactly two fields: "result" and "reason".
- "result" should be true or false (lowercase, JSON boolean). If you judge the entry corresponds to a real publication, output true; otherwise output false.
- "reason" should be a short reason (1-3 sentences).
- No extra keys, no markdown, no surrounding text.

Entry:
Cite Title: {title}
Authors: {authors}
Year: {year}
Venue: {venue}
\end{quote}

\subsection{Factors Affecting Hallucination Rates}

In this section, we provide additional analysis on factors that may influence citation hallucination rates in LLMs, including the impact of online search and chain-of-thought prompting, batch size effects, and domain-level sensitivity.

\paragraph{Online Search and Chain-of-Thought Impact}
\label{app:online_impact}

\Cref{tab:online_impact} reports citation hallucination rates with online search and chain-of-thought prompting versus baseline settings, along with per-model differences.
These settings are controlled through a third-party API aggregation interface, so results reflect the API-level toggles we could access rather than each vendor's full native retrieval or reasoning stack.

\begin{table}[t]
  \centering
  \begin{threeparttable}
    \caption{Citation hallucination rates: offline vs. online+thinking. $\Delta$ shows the difference between two modes.}
    \label{tab:online_impact}
    \small
\rowcolors{2}{tableRowLight}{tableRowWhite}
\begin{tabular}{lccc}
\toprule
\textbf{Model} & \cellcolor{red!12}\textbf{Offline (\%)} & \cellcolor{red!12}\textbf{Online (\%)} & \textbf{$\Delta$ (\%)} \\
\midrule
Seed & \cellcolor{red!30}75.78 & \cellcolor{red!34}94.50 & \cellcolor{red!18}+18.72 \\
GPT-5 & \cellcolor{red!20}42.54 & \cellcolor{red!24}58.97 & \cellcolor{red!16}+16.42 \\
Llama 4 & \cellcolor{red!20}44.22 & \cellcolor{red!20}47.50 & \cellcolor{red!10}+3.27 \\
Claude 4 & \cellcolor{red!12}21.43 & \cellcolor{red!12}22.25 & \cellcolor{red!8}+0.83 \\
GLM-4.5 & \cellcolor{red!12}21.20 & \cellcolor{red!12}21.29 & \cellcolor{red!6}+0.10 \\
Phi-4 & \cellcolor{red!34}87.47 & \cellcolor{red!34}87.48 & \cellcolor{red!6}+0.02 \\
Gemini & \cellcolor{red!24}59.67 & \cellcolor{red!24}59.26 & \cellcolor{green!6}-0.42 \\
Grok 4 & \cellcolor{red!30}80.26 & \cellcolor{red!30}79.70 & \cellcolor{green!6}-0.57 \\
Qwen-3 & \cellcolor{red!12}23.83 & \cellcolor{red!12}23.21 & \cellcolor{green!6}-0.62 \\
Hunyuan & \cellcolor{red!34}95.27 & \cellcolor{red!34}94.59 & \cellcolor{green!6}-0.68 \\
Kimi & \cellcolor{red!20}42.28 & \cellcolor{red!20}41.45 & \cellcolor{green!8}-0.84 \\
DeepSeek & \cellcolor{red!10}14.68 & \cellcolor{red!10}13.78 & \cellcolor{green!8}-0.89 \\
ERNIE & \cellcolor{red!30}75.77 & \cellcolor{red!28}68.12 & \cellcolor{green!10}-7.65 \\
\midrule
\textbf{Average} & \cellcolor{red!22}52.65 & \cellcolor{red!24}54.78 & \cellcolor{red!10}\textbf{+2.13} \\
\bottomrule
\end{tabular}
  \end{threeparttable}
\end{table}

\paragraph{Batch Size Impact on Hallucination Rates}

\label{app:batch_size}

Table~\ref{tab:batch_size} presents the detailed breakdown of citation hallucination rates by batch size (10, 20, and 30 citations per prompt) for all 13 models.
The results show no consistent correlation between batch size and hallucination rates, suggesting that hallucination reflects fundamental limitations in the models' bibliographic knowledge rather than output quantity constraints.

\subsection{Domain-Level Hallucination Rates}
\label{app:domain_sensitivity}

\begin{table}[t]
  \centering
  \begin{threeparttable}
    \caption{Citation stability scores across repeated runs. Sorted by valid citation stability (higher means more consistent).}
    \label{tab:citation_stability}
    \small
\begin{tabular}{lcccc}
\toprule
\textbf{Model} & \multicolumn{2}{c}{\cellcolor{green!14}\textbf{Valid Stability}} & \multicolumn{2}{c}{\cellcolor{red!14}\textbf{Halluc. Stability}} \\
\cmidrule(lr){2-3} \cmidrule(lr){4-5}
 & \cellcolor{green!12}Mean & \cellcolor{green!12}Std & \cellcolor{red!12}Mean & \cellcolor{red!12}Std \\
\midrule
\cellcolor{tableRowLight}DeepSeek & \cellcolor{green!24}0.577 & \cellcolor{green!24}0.094 & \cellcolor{red!18}0.226 & \cellcolor{red!18}0.216 \\
\cellcolor{tableRowWhite}Qwen-3 & \cellcolor{green!24}0.574 & \cellcolor{green!24}0.090 & \cellcolor{red!16}0.175 & \cellcolor{red!16}0.199 \\
\cellcolor{tableRowLight}Claude 4 & \cellcolor{green!24}0.571 & \cellcolor{green!24}0.187 & \cellcolor{red!16}0.181 & \cellcolor{red!16}0.187 \\
\cellcolor{tableRowWhite}Gemini & \cellcolor{green!22}0.552 & \cellcolor{green!22}0.156 & \cellcolor{red!12}0.069 & \cellcolor{red!12}0.083 \\
\cellcolor{tableRowLight}GLM-4.5 & \cellcolor{green!20}0.528 & \cellcolor{green!20}0.115 & \cellcolor{red!16}0.190 & \cellcolor{red!16}0.203 \\
\cellcolor{tableRowWhite}GPT-5 & \cellcolor{green!20}0.512 & \cellcolor{green!20}0.117 & \cellcolor{red!10}0.040 & \cellcolor{red!10}0.036 \\
\cellcolor{tableRowLight}Llama 4 & \cellcolor{green!18}0.469 & \cellcolor{green!18}0.150 & \cellcolor{red!12}0.097 & \cellcolor{red!12}0.070 \\
\cellcolor{tableRowWhite}Grok 4 & \cellcolor{green!14}0.380 & \cellcolor{green!14}0.194 & \cellcolor{red!10}0.035 & \cellcolor{red!10}0.027 \\
\cellcolor{tableRowLight}Hunyuan & \cellcolor{green!14}0.348 & \cellcolor{green!14}0.255 & \cellcolor{red!8}0.024 & \cellcolor{red!8}0.040 \\
\cellcolor{tableRowWhite}Kimi & \cellcolor{green!14}0.341 & \cellcolor{green!14}0.090 & \cellcolor{red!10}0.041 & \cellcolor{red!10}0.080 \\
\cellcolor{tableRowLight}ERNIE & \cellcolor{green!12}0.323 & \cellcolor{green!12}0.164 & \cellcolor{red!10}0.034 & \cellcolor{red!10}0.040 \\
\cellcolor{tableRowWhite}Seed & \cellcolor{green!10}0.246 & \cellcolor{green!10}0.191 & \cellcolor{red!10}0.047 & \cellcolor{red!10}0.091 \\
\cellcolor{tableRowLight}Phi-4 & \cellcolor{green!8}0.191 & \cellcolor{green!8}0.146 & \cellcolor{red!6}0.007 & \cellcolor{red!6}0.012 \\
\bottomrule
\end{tabular}
    \begin{tablenotes}
      \footnotesize
      \item \colorbox{red!28}{\textbf{Halluc. Stability:}} consistency of hallucinated citations across runs;
    \end{tablenotes}
  \end{threeparttable}
\end{table}

Table~\ref{tab:domain_sensitivity} presents citation hallucination rates by research domain (40 CS subfields), aggregated across all 13 models and ranked from highest to lowest.

\begin{table*}[t]
  \centering
  \begin{threeparttable}
    \caption{Citation hallucination rates (\%) by research domain (40 CS subfields), ranked from highest to lowest.}
    \label{tab:domain_sensitivity}
    \small
\rowcolors{2}{tableRowLight}{tableRowWhite}
\begin{tabular}{lrr@{\hskip 2em}lrr}
\toprule
\textbf{Domain} & \cellcolor{red!12}\textbf{Halluc.} & \textbf{95\% CI} & \textbf{Domain} & \cellcolor{red!12}\textbf{Halluc.} & \textbf{95\% CI} \\
\midrule
DL (Digital Libraries) & \cellcolor{red!28}80.19 & $\pm$3.46 & FL (Formal Languages) & \cellcolor{red!20}49.60 & $\pm$4.32 \\
OH (Other Hardware) & \cellcolor{red!28}75.38 & $\pm$3.73 & DM (Discrete Math) & \cellcolor{red!20}49.33 & $\pm$4.30 \\
NI (Network Info.) & \cellcolor{red!28}73.42 & $\pm$3.75 & PL (Prog. Languages) & \cellcolor{red!20}48.40 & $\pm$4.28 \\
ET (Emerg. Tech.) & \cellcolor{red!28}69.38 & $\pm$3.97 & OS (Operating Sys.) & \cellcolor{red!20}47.96 & $\pm$4.32 \\
GT (Game Theory) & \cellcolor{red!26}64.04 & $\pm$4.10 & SC (Sound) & \cellcolor{red!18}47.23 & $\pm$4.28 \\
SY (Symbolic Comp.) & \cellcolor{red!26}63.90 & $\pm$4.12 & NE (Networking) & \cellcolor{red!18}45.58 & $\pm$4.26 \\
CY (Computers \& Soc.) & \cellcolor{red!26}62.24 & $\pm$4.22 & SD (Software Dev.) & \cellcolor{red!18}45.51 & $\pm$4.34 \\
AR (Architecture) & \cellcolor{red!24}60.59 & $\pm$4.21 & NA (Num. Analysis) & \cellcolor{red!18}44.01 & $\pm$4.27 \\
CE (Comp. Eng.) & \cellcolor{red!24}60.58 & $\pm$4.22 & DC (Distributed Comp.) & \cellcolor{red!18}43.96 & $\pm$4.29 \\
DB (Databases) & \cellcolor{red!24}60.31 & $\pm$4.23 & IT (Info. Theory) & \cellcolor{red!18}43.18 & $\pm$4.23 \\
SE (Software Eng.) & \cellcolor{red!24}60.26 & $\pm$4.19 & LG (Learning) & \cellcolor{red!16}41.92 & $\pm$4.30 \\
CR (Cryptography) & \cellcolor{red!24}59.96 & $\pm$4.20 & LO (Logic in CS) & \cellcolor{red!16}41.55 & $\pm$4.28 \\
HC (Human-Comp. Int.) & \cellcolor{red!24}59.51 & $\pm$4.27 & MS (Math. Software) & \cellcolor{red!14}35.03 & $\pm$4.11 \\
MM (Multimedia) & \cellcolor{red!24}59.20 & $\pm$4.23 & CV (Comp. Vision) & \cellcolor{red!14}33.64 & $\pm$4.08 \\
AI (Artificial Intel.) & \cellcolor{red!22}57.32 & $\pm$4.23 & GR (Graphics) & \cellcolor{red!14}32.50 & $\pm$4.07 \\
PF (Performance) & \cellcolor{red!22}54.63 & $\pm$4.31 & CC (Comp. Complexity) & \cellcolor{red!14}32.23 & $\pm$4.02 \\
MA (Multiagent Sys.) & \cellcolor{red!22}53.97 & $\pm$4.30 & GL (General Lit.) & \cellcolor{red!12}31.50 & $\pm$4.03 \\
RO (Robotics) & \cellcolor{red!22}52.21 & $\pm$4.37 & DS (Data Structures) & \cellcolor{red!12}29.76 & $\pm$3.93 \\
IR (Info. Retrieval) & \cellcolor{red!20}51.05 & $\pm$4.29 & SI (Social/Info. Net.) & \cellcolor{red!12}29.31 & $\pm$3.92 \\
CG (Comp. Geometry) & \cellcolor{red!20}49.66 & $\pm$4.31 & CL (Comp. Linguistics) & \cellcolor{red!12}28.80 & $\pm$4.00 \\
\bottomrule
\end{tabular}
    \begin{tablenotes}
      \footnotesize
      \item \colorbox{red!15}{\textbf{Halluc.:}} Hallucination rate, the percentage of citations verified as invalid.
    \end{tablenotes}
  \end{threeparttable}
\end{table*}

\subsection{Citation Stability and Cross-Model Overlap}

\label{app:stability_overlap}
\Cref{tab:citation_stability} reports stability statistics for valid versus hallucinated citations.
We define \emph{citation stability} as the degree to which a model generates identical citations when repeatedly prompted with the same configuration.
Concretely, for each (model, config) pair, where a config is defined by the combination of research topic, requested number of references, chain-of-thought setting, and online-search setting. We aggregate all citations produced across repeated runs.
The stability score is computed as:
\[
\text{Stability} = 1 - \frac{|\text{Unique Titles}|}{|\text{Total Citations}|}
\]
A score of 1 indicates perfect consistency (the model always outputs the same papers), while a score of 0 indicates maximal variability (no repetition across runs).
As shown in \cref{tab:citation_stability}, valid citations exhibit markedly higher stability (mean up to 0.58 for DeepSeek and 0.57 for Qwen-3) than hallucinated ones (mean up to 0.23).
This suggests that real papers that are strongly encoded in parametric memory are recalled repeatedly, whereas fabricated references are generated more arbitrarily.

Table~\ref{tab:most_cited_by_topic} presents the most frequently generated valid citations for each of the 40 research domains in our LLM benchmark experiment.
These papers represent the references that LLMs most consistently recall, suggesting they are strongly encoded in the models' parametric memory due to high frequency in training data.
Seminal works such as ``NeRF'' (Graphics), ``RAG'' (Computation \& Language), and ``U-Net'' (Computer Vision) appear at the top, reflecting the influence of foundational papers in each field.

\begin{table*}[t]
  \centering
  \begin{threeparttable}
    \caption{Most frequently generated valid citations by research domain across all 13 LLMs.}
    \label{tab:most_cited_by_topic}
    \small
\rowcolors{2}{tableRowLight}{tableRowWhite}
\begin{tabular}{clrc}
\toprule
\textbf{Topic} & \textbf{Most Cited Paper Title} & \textbf{Count} & \textbf{\%} \\
\midrule
GR & NeRF: Representing Scenes as Neural Radiance Fields for View Synthesis & \cellcolor{green!28}352 & \cellcolor{green!28}6.46 \\
CL & Retrieval-Augmented Generation for Knowledge-Intensive NLP Tasks & \cellcolor{green!28}348 & \cellcolor{green!28}6.24 \\
NI & Green Traffic Engineering for Satellite Networks Using Segment Routing Flexible Algorithm & \cellcolor{green!24}138 & \cellcolor{green!24}5.97 \\
ET & FCT O-RAN: Design and Deployment of a Multi-Vendor End-to-End Private 5G Testbed & \cellcolor{green!24}148 & \cellcolor{green!24}5.69 \\
AR & Eyeriss: An Energy-Efficient Reconfigurable Accelerator for Deep Convolutional Neural Networks & \cellcolor{green!24}184 & \cellcolor{green!24}5.55 \\
DL & Digital Libraries & \cellcolor{green!24}96 & \cellcolor{green!24}5.55 \\
IT & A Mathematical Theory of Communication & \cellcolor{green!24}260 & \cellcolor{green!24}5.46 \\
SI & Semi-Supervised Classification with Graph Convolutional Networks & \cellcolor{green!16}287 & \cellcolor{green!16}4.88 \\
CV & U-Net: Convolutional Networks for Biomedical Image Segmentation & \cellcolor{green!16}262 & \cellcolor{green!16}4.83 \\
GT & Algorithmic Game Theory & \cellcolor{green!16}146 & \cellcolor{green!16}4.62 \\
DC & MapReduce: Simplified Data Processing on Large Clusters & \cellcolor{green!16}202 & \cellcolor{green!16}4.47 \\
AI & Language Models are Few-Shot Learners & \cellcolor{green!16}154 & \cellcolor{green!16}4.28 \\
SD & Musical genre classification of audio signals & \cellcolor{green!16}185 & \cellcolor{green!16}4.27 \\
CG & Computational Geometry: Algorithms and Applications & \cellcolor{green!16}178 & \cellcolor{green!16}4.19 \\
SE & Automated Generation of Issue-Reproducing Tests by Combining LLMs and Search-Based Testing & \cellcolor{green!16}141 & \cellcolor{green!16}4.14 \\
FL & Introduction to Automata Theory, Languages, and Computation & \cellcolor{green!16}171 & \cellcolor{green!16}4.10 \\
GL & Denoising Diffusion Probabilistic Models & \cellcolor{green!16}234 & \cellcolor{green!16}4.08 \\
CC & Computational Complexity: A Modern Approach & \cellcolor{green!16}230 & \cellcolor{green!16}4.07 \\
HC & AttenTrack: Mobile User Attention Awareness Based on Context and External Distractions & \cellcolor{green!12}130 & \cellcolor{green!12}3.93 \\
MA & Multiagent Systems: Algorithmic, Game-Theoretic, and Logical Foundations & \cellcolor{green!12}152 & \cellcolor{green!12}3.89 \\
MM & LLM-Guided Semantic Relational Reasoning for Multimodal Intent Recognition & \cellcolor{green!12}134 & \cellcolor{green!12}3.83 \\
PL & Dato: A Task-Based Programming Model for Dataflow Accelerators & \cellcolor{green!12}168 & \cellcolor{green!12}3.80 \\
CY & Algorithms of Oppression: How Search Engines Reinforce Racism & \cellcolor{green!12}114 & \cellcolor{green!12}3.56 \\
OS & Scheduling Algorithms for Multiprogramming in a Hard-Real-Time Environment & \cellcolor{green!12}153 & \cellcolor{green!12}3.51 \\
DS & Introduction to Algorithms & \cellcolor{green!12}202 & \cellcolor{green!12}3.42 \\
OH & Cutting the Electric Bill for Internet-Scale Systems & \cellcolor{green!12}71 & \cellcolor{green!12}3.34 \\
MS & Robust Regression and Outlier Detection & \cellcolor{green!12}180 & \cellcolor{green!12}3.25 \\
PF & Optimal Parallel Scheduling under Concave Speedup Functions & \cellcolor{green!12}122 & \cellcolor{green!12}3.25 \\
CE & A continuum multi-species biofilm model with a novel interaction scheme & \cellcolor{green!12}103 & \cellcolor{green!12}3.15 \\
LO & Language-Based Information-Flow Security & \cellcolor{green!12}145 & \cellcolor{green!12}3.03 \\
LG & Causality: Models, Reasoning, and Inference & \cellcolor{green!8}140 & \cellcolor{green!8}2.92 \\
SY & Three-Phase PLLs: A Review of Recent Advances & \cellcolor{green!8}81 & \cellcolor{green!8}2.66 \\
IR & BERT4Rec: Sequential Recommendation with Bidirectional Encoder Representations from Transformer & \cellcolor{green!8}110 & \cellcolor{green!8}2.65 \\
CR & A Method for Obtaining Digital Signatures and Public-Key Cryptosystems & \cellcolor{green!8}91 & \cellcolor{green!8}2.61 \\
NA & Numerical Linear Algebra & \cellcolor{green!8}121 & \cellcolor{green!8}2.56 \\
DM & Polyphase codes with good periodic correlation properties & \cellcolor{green!8}107 & \cellcolor{green!8}2.50 \\
SC & Planning Algorithms & \cellcolor{green!8}110 & \cellcolor{green!8}2.45 \\
RO & Human-Robot Interaction: A Survey & \cellcolor{green!8}96 & \cellcolor{green!8}2.44 \\
NE & Handbook of Evolutionary Computation & \cellcolor{green!8}114 & \cellcolor{green!8}2.42 \\
DB & MapReduce: Simplified Data Processing on Large Clusters & \cellcolor{green!8}71 & \cellcolor{green!8}2.08 \\
\bottomrule
\end{tabular}
  \end{threeparttable}
\end{table*}

\subsection{Survey Response Details (Full)}
\label{app:survey_details_full}

\cref{tab:survey_details_full_a1,tab:survey_details_full_a2,tab:survey_details_full_b1,tab:survey_details_full_b2} provide the complete numerical breakdown of all survey responses (N=94 valid responses, 97 total), including every response option for adoption, verification, severity, responsibility, support for automated tools, risky behaviors, reviewer reactions, text polishing, primary metadata source, and claim verification frequency. The question wording and numbering map is shown in \cref{tab:survey_question_map}.


\begin{table*}[htbp]
\centering
\caption{Survey question numbering map.}
\label{tab:survey_question_map}
\small
\begin{threeparttable}
\begin{tabular}{p{1.2cm}p{14.2cm}}
\toprule
\textbf{Q\#} & \textbf{Question} \\
\midrule
Q1 & What is your current academic position? \\
Q2 & Which research area best describes your work? \\
Q3 & How many peer-reviewed papers have you published in your career? \\
Q4 & Have you served as a reviewer for top-tier conferences (e.g., USENIX Sec, CCS, S\&P, CVPR, NeurIPS) in the last 3 years? \\
Q5 & Which stage of paper writing do you find most time-consuming? (Select up to 2) \\
Q6 & How do you typically search for related work? \\
Q7 & Do you use AI-powered tools for research? \\
Q8 & Which types of AI-powered tools do you use for research? (Select all that apply) \\
Q9 & Do you use AI tools specifically for text polishing, grammar checking, or rephrasing? \\
Q10 & Specifically regarding citations, how do you use AI tools? (Select all that apply) \\
Q11 & When asking a general LLM (e.g., ChatGPT) to find references, how often do you encounter hallucinations (non-existent papers)? \\
Q12 & If an AI tool provides a perfect-looking reference (Title, Author, Year, Venue all look correct), do you still verify it externally? \\
Q13 & Have you ever cited a paper suggested by AI without reading the full text of that paper? \\
Q14 & To what extent do you agree: AI tools have made it easier to generate related work sections, but harder to ensure citation accuracy. \\
Q15 & What is your strategy when an AI tool generates a citation with a broken link or DOI? \\
Q16 & When adding a citation to your paper, what is your primary source of metadata (title, year, venue)? \\
Q17 & How often do you verify that a cited paper actually contains the claim you are attributing to it? \\
Q18 & If you cannot find the full text of a paper (e.g., paywalled or offline), do you still cite it based on its abstract or title? \\
Q19 & To what extent do you agree: There is pressure to include a high number of references to make the paper look more scholarly. \\
Q20 & When reviewing a paper, do you explicitly check the Reference section? \\
Q21 & What are you looking for when checking references? (Select all that apply) \\
Q22 & Have you ever clicked on a DOI link in a submission's bibliography to verify the paper exists? \\
Q23 & Have you ever suspected a submission contained fake or hallucinated references? \\
Q24 & If you see a citation to a preprint or ArXiv paper, do you verify if it has been published in a peer-reviewed venue? \\
Q25 & How do you handle citations to non-English papers in a submission? \\
Q26 & To what extent do you agree: The accuracy of the bibliography is as important as the accuracy of the experimental results. \\
Q27 & Do you believe the current peer review process is effective at catching metadata errors in references? \\
Q28 & How serious of an issue do you consider hallucinated citations (non-existent papers) in the era of AI? \\
Q29 & Who is primarily responsible for verifying the authenticity of cited references? \\
Q30 & Should conferences deploy automated tools to check for broken DOIs or fake references upon submission? \\
Q31 & To what extent do you agree: It is acceptable to cite a paper based on AI summarization without reading the original text. \\
Q32 & Have you ever seen a reference list where the author names were clearly scrambled or incorrect (e.g., order reversed)? \\
Q33 & Have you ever seen a reference where the title existed but the venue/year was completely wrong? \\
Q34 & Do you think it is acceptable to cite a technical report if a peer-reviewed version exists? \\
Q35 & What is your reaction if you find one incorrect citation (e.g., wrong year) in a paper you are reviewing? \\
Q36 & What is your reaction if you find multiple (e.g., >5) incorrect or fake citations? \\
Q37 & If you encounter a suspicious or clearly fake reference in a published paper, do you report it? \\
Q38$^*$ & Have you ever copy-pasted a BibTeX entry from the internet without checking its content? \\
Q39$^*$ & To what extent do you agree: I meticulously verify every single field (volume, issue, page numbers) of every BibTeX entry I import, ensuring 100\% accuracy before submission. \\
Q40 & Do you use tools like Semantic Scholar or ResearchRabbit to build your bibliography? \\
\bottomrule
\end{tabular}
\begin{tablenotes}[para,flushleft]
    \footnotesize
    \textbf{Reverse-worded consistency check:} Q38 ``risky behavior'' vs.~Q39 ``diligent behavior.'' Three respondents answered inconsistently (``often copy-paste'' AND ``strongly verify'') and were excluded.
\end{tablenotes}
\end{threeparttable}
\end{table*}

\begin{table*}[htbp]
  \centering
  \begin{threeparttable}
    \caption{Full survey response details, Part A-1: Q1--Q10 (N=94 valid responses).}
    \label{tab:survey_details_full_a1}
    \small
    \begin{tabular}{p{6cm}rr|p{6cm}rr}
\toprule
\textbf{Item} & \textbf{N} & \textbf{\%} & \textbf{Item} & \textbf{N} & \textbf{\%} \\
\midrule
\textit{Q1: What is your current academic position?} & & & \textit{Q2: Which research area best describes your work?} & & \\
\rowcolor{tableRowLight}
PhD Student & 32 & 34.0 & Other & 24 & 21.1 \\
Master's Student & 19 & 20.2 & AI & 20 & 17.5 \\
\rowcolor{tableRowLight}
Faculty (Professor/Lecturer) & 17 & 18.1 & Machine Learning & 20 & 17.5 \\
Undergraduate student & 13 & 13.8 & Network Security & 18 & 15.8 \\
\rowcolor{tableRowLight}
Postdoc & 6 & 6.4 & AI Security & 13 & 11.4 \\
Researcher & 5 & 5.3 & System Security & 10 & 8.8 \\
\rowcolor{tableRowLight}
Other & 2 & 2.1 & Cryptography & 5 & 4.4 \\
 &  &  & Software Engineering & 4 & 3.5 \\
\midrule
\textit{Q3: How many peer-reviewed papers have you published in your career?} & & & \textit{Q4: Have you served as a reviewer for top-tier conferences (e.g., USENIX Sec, CCS, S\&P, CVPR, NeurIPS) in the last 3 years?} & & \\
\rowcolor{tableRowLight}
1--5 & 46 & 48.9 & No & 62 & 66.0 \\
6--10 & 16 & 17.0 & Yes & 32 & 34.0 \\
\rowcolor{tableRowLight}
0 & 14 & 14.9 &  &  &  \\
20+ & 12 & 12.8 &  &  &  \\
\rowcolor{tableRowLight}
11--20 & 6 & 6.4 &  &  &  \\
\midrule
\textit{Q5: Which stage of paper writing do you find most time-consuming? (Select up to 2)} & & & \textit{Q6: How do you typically search for related work?} & & \\
\rowcolor{tableRowLight}
Conceptualization & 62 & 35.4 & Keyword search on Google Scholar/DBLP & 68 & 33.5 \\
Writing the Introduction & 36 & 20.6 & Following citations from other papers & 62 & 30.5 \\
\rowcolor{tableRowLight}
Literature Search & 24 & 13.7 & Browsing conference proceedings & 40 & 19.7 \\
Review & 24 & 13.7 & Using AI-powered tools (e.g., ChatGPT, Connected Papers) & 33 & 16.3 \\
\rowcolor{tableRowLight}
Technical Description & 24 & 13.7 &  &  &  \\
Formatting References & 5 & 2.9 &  &  &  \\
\midrule
\textit{Q7: Do you use AI-powered tools for research?} & & & \textit{Q8: Which types of AI-powered tools do you use for research? (Select all that apply)} & & \\
\rowcolor{tableRowLight}
Yes & 75 & 87.2 & General-purpose LLMs (ChatGPT, Claude, Gemini) & 75 & 58.6 \\
No & 11 & 12.8 & Coding Assistants (Copilot, Cursor) & 41 & 32.0 \\
\rowcolor{tableRowLight}
 &  &  & Academic search (Elicit, Semantic Scholar) & 6 & 4.7 \\
 &  &  & AI reading tools (ChatPDF, Humata) & 6 & 4.7 \\
\midrule
\textit{Q9: Do you use AI tools specifically for text polishing, grammar checking, or rephrasing?} & & & \textit{Q10: Specifically regarding citations, how do you use AI tools? (Select all that apply)} & & \\
\rowcolor{tableRowLight}
Yes, for specific difficult paragraphs & 35 & 46.7 & None (do not use AI for citation tasks) & 35 & 29.9 \\
Yes, for almost every sentence & 22 & 29.3 & Formatting (convert to BibTeX, etc.) & 23 & 19.7 \\
\rowcolor{tableRowLight}
Yes, only for final proofreading & 17 & 22.7 & Discovery (recommend papers on topic) & 23 & 19.7 \\
No, trust my own writing more & 1 & 1.3 & Summarization (decide whether to cite) & 15 & 12.8 \\
\rowcolor{tableRowLight}
 &  &  & Gap-filling (need a citation) & 15 & 12.8 \\
 &  &  & Verification (ask if paper exists) & 6 & 5.1 \\
\bottomrule
    \end{tabular}
  \end{threeparttable}
\end{table*}

\begin{table*}[htbp]
  \centering
  \begin{threeparttable}
    \caption{Full survey response details, Part A-2: Q11--Q20 (N=94 valid responses).}
    \label{tab:survey_details_full_a2}
    \small
    \begin{tabular}{p{6cm}rr|p{6cm}rr}
\toprule
\textbf{Item} & \textbf{N} & \textbf{\%} & \textbf{Item} & \textbf{N} & \textbf{\%} \\
\midrule
\textit{Q11: When asking a general LLM (e.g., ChatGPT) to find references, how often do you encounter hallucinations (non-existent papers)?} & & & \textit{Q12: If an AI tool provides a perfect-looking reference (Title, Author, Year, Venue all look correct), do you still verify it externally?} & & \\
\rowcolor{tableRowLight}
Often (20--50\%) & 24 & 32.0 & Always (e.g., Scholar/DBLP) & 58 & 77.3 \\
Don't use LLMs for finding refs & 25 & 33.3 & Only if reading full text & 8 & 10.7 \\
\rowcolor{tableRowLight}
Occasionally (<20\%) & 16 & 21.3 & Yes, verify 100\% via Scholar/DBLP & 7 & 9.3 \\
Very Often (>50\%) & 7 & 9.3 & Only if suspicious & 1 & 1.3 \\
\rowcolor{tableRowLight}
Never & 3 & 4.0 & Only if need to read the full text & 1 & 1.3 \\
\midrule
\textit{Q13: Have you ever cited a paper suggested by AI without reading the full text of that paper?} & & & \textit{Q14: To what extent do you agree: AI tools have made it easier to generate related work sections, but harder to ensure citation accuracy.} & & \\
\rowcolor{tableRowLight}
No, never & 62 & 82.7 & Somewhat agree & 28 & 37.3 \\
Yes, once or twice & 8 & 10.7 & Strongly agree & 24 & 32.0 \\
\rowcolor{tableRowLight}
Yes, often & 5 & 6.7 & Neutral & 11 & 14.7 \\
 &  &  & Somewhat disagree & 9 & 12.0 \\
\rowcolor{tableRowLight}
 &  &  & Strongly disagree & 3 & 4.0 \\
\midrule
\textit{Q15: What is your strategy when an AI tool generates a citation with a broken link or DOI?} & & & \textit{Q16: When adding a citation to your paper, what is your primary source of metadata (title, year, venue)?} & & \\
\rowcolor{tableRowLight}
Assume exists, find manually & 45 & 60.0 & Google Scholar ``Cite'' button & 61 & 72.6 \\
Assume hallucination, discard & 24 & 32.0 & Direct export from publisher & 15 & 17.9 \\
\rowcolor{tableRowLight}
Ask AI for different link & 4 & 5.3 & Direct export from publisher (IEEE/ACM) & 4 & 4.8 \\
\rowcolor{tableRowLight}
Keep citation, remove DOI & 2 & 2.7 &  &  &  \\
\midrule
\textit{Q17: How often do you verify that a cited paper actually contains the claim you are attributing to it?} & & & \textit{Q18: If you cannot find the full text of a paper (e.g., paywalled or offline), do you still cite it based on its abstract or title?} & & \\
\rowcolor{tableRowLight}
Every single time & 48 & 57.1 & No, never & 46 & 54.8 \\
Most of the time & 20 & 23.8 & Yes, if necessary & 35 & 41.7 \\
\rowcolor{tableRowLight}
Only for critical claims & 14 & 16.7 & Yes, often & 3 & 3.6 \\
Rarely & 2 & 2.4 &  &  &  \\
\midrule
\textit{Q19: To what extent do you agree: There is pressure to include a high number of references to make the paper look more scholarly.} & & & \textit{Q20: When reviewing a paper, do you explicitly check the Reference section?} & & \\
\rowcolor{tableRowLight}
Somewhat agree & 31 & 36.9 & Yes, skimming & 18 & 60.0 \\
Neutral & 21 & 25.0 & Yes, carefully & 7 & 23.3 \\
\rowcolor{tableRowLight}
Somewhat disagree & 17 & 20.2 & No & 5 & 16.7 \\
Strongly disagree & 10 & 11.9 &  &  &  \\
\rowcolor{tableRowLight}
Strongly agree & 5 & 6.0 &  &  &  \\
\bottomrule
    \end{tabular}
  \end{threeparttable}
\end{table*}

\begin{table*}[htbp]
  \centering
  \begin{threeparttable}
    \caption{Full survey response details, Part B-1: Q21--Q30 (N=94 valid responses).}
    \label{tab:survey_details_full_b1}
    \small
    \begin{tabular}{p{6cm}rr|p{6cm}rr}
\toprule
\textbf{Item} & \textbf{N} & \textbf{\%} & \textbf{Item} & \textbf{N} & \textbf{\%} \\
\midrule
\textit{Q21: What are you looking for when checking references? (Select all that apply)} & & & \textit{Q22: Have you ever clicked on a DOI link in a submission's bibliography to verify the paper exists?} & & \\
\rowcolor{tableRowLight}
Missing key related work & 28 & 52.8 & Occasionally & 13 & 43.3 \\
Correctness of metadata & 10 & 18.9 & Rarely & 9 & 30.0 \\
\rowcolor{tableRowLight}
Existence of papers & 10 & 18.9 & Never & 6 & 20.0 \\
Self-citation abuse & 5 & 9.4 & Frequently & 2 & 6.7 \\
\midrule
\textit{Q23: Have you ever suspected a submission contained fake or hallucinated references?} & & & \textit{Q24: If you see a citation to a preprint or ArXiv paper, do you verify if it has been published in a peer-reviewed venue?} & & \\
\rowcolor{tableRowLight}
No & 24 & 80.0 & Sometimes & 15 & 50.0 \\
Yes & 6 & 20.0 & No & 9 & 30.0 \\
\rowcolor{tableRowLight}
 &  &  & Always & 6 & 20.0 \\
\midrule
\textit{Q25: How do you handle citations to non-English papers in a submission?} & & & \textit{Q26: To what extent do you agree: The accuracy of the bibliography is as important as the accuracy of the experimental results.} & & \\
\rowcolor{tableRowLight}
I check them using translation tools & 10 & 33.3 & Strongly agree & 52 & 55.3 \\
I ignore them & 10 & 33.3 & Somewhat agree & 29 & 30.9 \\
\rowcolor{tableRowLight}
I assume they are valid & 9 & 30.0 & Neutral & 10 & 10.6 \\
I ask the authors to clarify & 1 & 3.3 & Somewhat disagree & 3 & 3.2 \\
\rowcolor{tableRowLight}
\midrule
\textit{Q27: Do you believe the current peer review process is effective at catching metadata errors in references?} & & & \textit{Q28: How serious of an issue do you consider hallucinated citations (non-existent papers) in the era of AI?} & & \\
\rowcolor{tableRowLight}
Not very effective & 62 & 66.0 & Critical crisis & 42 & 44.7 \\
Somewhat effective & 22 & 23.4 & Major problem & 30 & 31.9 \\
\rowcolor{tableRowLight}
Ineffective & 8 & 8.5 & Minor nuisance & 22 & 23.4 \\
Very effective & 2 & 2.1 &  &  &  \\
\midrule
\textit{Q29: Who is primarily responsible for verifying the authenticity of cited references?} & & & \textit{Q30: Should conferences deploy automated tools to check for broken DOIs or fake references upon submission?} & & \\
\rowcolor{tableRowLight}
Authors & 86 & 91.5 & Yes, absolutely & 66 & 70.2 \\
Reviewers & 3 & 3.2 & Maybe & 25 & 26.6 \\
\rowcolor{tableRowLight}
Publishers (Editorial) & 1 & 1.1 & No & 3 & 3.2 \\
AI tool developers & 2 & 2.1 &  &  &  \\
\bottomrule
    \end{tabular}
  \end{threeparttable}
\end{table*}

\begin{table*}[htbp]
  \centering
  \begin{threeparttable}
    \caption{Full survey response details, Part B-2: Q31--Q40 (N=94 valid responses).}
    \label{tab:survey_details_full_b2}
    \small
    \begin{tabular}{p{6cm}rr|p{6cm}rr}
\toprule
\textbf{Item} & \textbf{N} & \textbf{\%} & \textbf{Item} & \textbf{N} & \textbf{\%} \\
\midrule
\textit{Q31: To what extent do you agree: It is acceptable to cite a paper based on AI summarization without reading the original text.} & & & \textit{Q32: Have you ever seen a reference list where the author names were clearly scrambled or incorrect (e.g., order reversed)?} & & \\
\rowcolor{tableRowLight}
Strongly disagree & 33 & 35.1 & No & 59 & 62.8 \\
Somewhat disagree & 31 & 33.0 & Yes & 35 & 37.2 \\
\rowcolor{tableRowLight}
Neutral & 17 & 18.1 &  &  &  \\
Somewhat agree & 12 & 12.8 &  &  &  \\
\rowcolor{tableRowLight}
Strongly agree & 1 & 1.1 &  &  &  \\
\midrule
\textit{Q33: Have you ever seen a reference where the title existed but the venue/year was completely wrong?} & & & \textit{Q34: Do you think it is acceptable to cite a technical report if a peer-reviewed version exists?} & & \\
\rowcolor{tableRowLight}
No & 50 & 53.2 & Yes & 72 & 76.6 \\
Yes & 44 & 46.8 & No & 22 & 23.4 \\
\midrule
\textit{Q35: What is your reaction if you find one incorrect citation (e.g., wrong year) in a paper you are reviewing?} & & & \textit{Q36: What is your reaction if you find multiple (e.g., >5) incorrect or fake citations?} & & \\
\rowcolor{tableRowLight}
Mention in minor comments & 61 & 64.9 & Reject immediately (ethical concern) & 56 & 59.6 \\
Ask for major revision & 15 & 16.0 & Ask for explanation & 38 & 40.4 \\
\rowcolor{tableRowLight}
Reject the paper & 4 & 4.3 &  &  &  \\
\midrule
\textit{Q37: If you encounter a suspicious or clearly fake reference in a published paper, do you report it?} & & & \textit{Q38: Have you ever copy-pasted a BibTeX entry from the internet without checking its content?}$^*$ & & \\
\rowcolor{tableRowLight}
Verify privately, take no action & 40 & 37.0 & No, never & 55 & 58.5 \\
Contact PC Chairs / Journal Editors & 36 & 33.3 & Yes, rarely & 26 & 27.7 \\
\rowcolor{tableRowLight}
Contact authors directly & 24 & 22.2 & Yes, often & 13 & 13.8 \\
Ignore & 8 & 7.4 &  &  &  \\
\midrule
\textit{Q39: To what extent do you agree: I meticulously verify every single field (volume, issue, page numbers) of every BibTeX entry I import, ensuring 100\% accuracy before submission.}$^*$ & & & \textit{Q40: Do you use tools like Semantic Scholar or ResearchRabbit to build your bibliography?} & & \\
\rowcolor{tableRowLight}
Somewhat agree & 27 & 28.7 & No & 78 & 83.0 \\
Strongly agree & 26 & 27.7 & Yes & 16 & 17.0 \\
\rowcolor{tableRowLight}
Neutral & 26 & 27.7 &  &  &  \\
Somewhat disagree & 10 & 10.6 &  &  &  \\
\rowcolor{tableRowLight}
Strongly disagree & 5 & 5.3 &  &  &  \\
\bottomrule
    \end{tabular}
    \begin{tablenotes}[para,flushleft]
      \footnotesize
      \item Q38 and Q39 are semantically opposite questions. Q38 measures risky behavior (``Have you ever copy-pasted a BibTeX entry from the internet without checking its content?''), while Q39 measures diligent behavior (``To what extent do you agree: I meticulously verify every single field (volume, issue, page numbers) of every BibTeX entry I import, ensuring 100\% accuracy before submission.'').
    \end{tablenotes}
  \end{threeparttable}
\end{table*}

\end{document}